%% file: sample631.tex
\documentclass{aastex631}

\newcommand{\ir}[1]{#1}
\newcommand{\irtwo}[1]{#1}
\newcommand{\cwr}[1]{#1}

\usepackage{hyperref}
\usepackage[margin=1in]{geometry}  

\newcommand\vtpf{\texttt{LSST v3.4 - baseline}}
\newcommand\vtpfur{\texttt{LSST v3.4 - uniform rolling}}
\newcommand\vfpz{\texttt{LSST v4.0 - Phase 3 baseline}}
\newcommand\vfpzfc{\texttt{LSST v4.0 - four cycles}}
\newcommand\noroll{\texttt{LSST v3.4  - no rolling}}

\defcitealias{scoc_phase1}{Rubin SCOC Phase 1 Report}
\defcitealias{scoc_phase2}{Rubin SCOC Phase 2 Report}
\defcitealias{scoc_phase3}{Rubin SCOC Phase 3 Report}

\begin{document}

    \title{{Uniform Rolling:  An LSST \cwr{Observing Cadence} Offering Sufficient \cwr{Survey} Uniformity for \cwr{Comprehensive} Cosmological Analys\cwr{i}s}}

\author[0000-0002-3962-9274]{Boris Leistedt}
\affiliation{Department of Physics, Imperial College London, Blackett Laboratory, Prince Consort
Road, London SW7 2AZ, UK}

\author[0000-0001-7774-2246]{Matthew R. Becker}
\affiliation{High Energy Physics Division, Argonne National Laboratory, Lemont, IL 60439, USA}

\author[0000-0003-2296-7717]{Humna Awan}
\affiliation{Kavli Institute for Particle Astrophysics and Cosmology, Department of Physics, Stanford University, Stanford, CA 94309, USA}
\affiliation{SLAC National Accelerator Laboratory, 2575 Sand Hill Road, Menlo Park, CA 94025, USA}

\author[0000-0003-1530-8713]{Eric Gawiser}
\affiliation{Department of Physics and Astronomy, Rutgers, the State University of New Jersey, Piscataway, NJ 08854, USA}

\author{Qianjun Hang}
\affiliation{Department of Physics \& Astronomy, University College London, Gower Street, London WC1E 6BT, UK}

\author[0000-0002-9436-8871]{Ren\'ee Hlo\v{z}ek}
\affiliation{Department of Astronomy and Astrophysics, University of Toronto, 50 St. George Street, Toronto, ON M5S 3H4, Canada}
\affiliation{Dunlap Institute for Astronomy and Astrophysics, University of Toronto, 50 St. George Street, Toronto, ON M5S 3H4, Canada}

\author[0000-0001-8738-6011]{Saurabh W.~Jha}
\affiliation{Department of Physics and Astronomy, Rutgers, the State University of New Jersey, Piscataway, NJ 08854, USA}

\author[0000-0001-5916-0031]{R.~Lynne~Jones}
\affiliation{Vera C.~Rubin Observatory, 950 N. Cherry Ave., Tucson, AZ 85719, USA}

\author[0000-0001-8783-6529]{Arun Kannawadi}
\affiliation{Department of Physics, Duke University, Durham, NC 27708, USA}

\author[0000-0003-2221-8281]{Michelle Lochner}
\affiliation{Department of Physics and Astronomy, University of the Western Cape, Bellville, Cape Town, 7535, South Africa}

\author[0000-0003-2271-1527]{Rachel Mandelbaum}
\affiliation{McWilliams Center for Cosmology and Astrophysics, Department of Physics, Carnegie Mellon University, Pittsburgh, PA 15213, USA}

\author[0000-0001-8684-2222]{Jeffrey A. Newman}
\affiliation{Department of Physics \& Astronomy and Pittsburgh Particle Physics, Astrophysics, and Cosmology Center(PITTPACC), University of Pittsburgh, 3941O HaraStreet, Pittsburgh, PA 15260, USA}

\author[0000-0002-1831-1953]{I.~Sevilla-Noarbe}
\affiliation{Centro de Investigaciones Energ\'eticas, Medioambientales y Tecnol\'ogicas (CIEMAT), Madrid, Spain}

\author[0000-0002-2519-584X]{Hiranya V.\ Peiris}
\affiliation{Institute of Astronomy and Kavli Institute for Cosmology, University of Cambridge, Madingley Road, Cambridge CB3 0HA, UK}
\affiliation{The Oskar Klein Centre, Department of Physics, Stockholm University, AlbaNova University Centre, SE 106 91 Stockholm, Sweden}

\author[0000-0001-9376-3135]{Eli S.~Rykoff}
\affiliation{SLAC National Accelerator Laboratory, 2575 Sand Hill Road, Menlo Park, CA 94025, USA}

\author[0000-0002-5622-5212]{M.~A.~Troxel}
\affiliation{Department of Physics, Duke University, Durham, NC 27708, USA}

\author[0000-0003-2874-6464]{Peter Yoachim}
\affiliation{Department of Astronomy and the DiRAC Institute, University of Washington, Seattle, WA 98195, USA}

\collaboration{20}{The LSST Dark Energy Science Collaboration}

\begin{abstract}
The \cwr{Legacy Survey of Space and Time (LSST) that will be carried out by the NSF-DOE} Vera C.~Rubin Observatory promises to
be the defining survey of the next decade, supplying unprecedented access to the night sky
to static science- and time-domain science-focused researchers alike. Maximizing the output of
the broad remit of Rubin Observatory science requires a non-trivial \textit{survey strategy} (i.e.,
the sequence of observations in space, time, and passband).  For time-domain science,
the most promising strategy designed so far is a \textit{rolling} survey strategy, whereby a subset of the
full LSST survey area is observed at higher rate compared with the nominal rate
dictated by weather conditions and the \cwr{observatory's} technical constraints. 
This strategy is now the baseline approach for
the LSST as a whole. Focusing on static science (galaxy clustering and weak lensing), we study how these time-domain-optimized rolling strategies
affect the \ir{depth uniformity at intermediate years of the survey}. 
We characterize the amount of survey area at high risk of
being lost in static-science analyses of a baseline rolling LSST dataset
due to an insufficient combination of survey contiguity and uniformity. At intermediate data releases,
nearly half of the survey could be lost for static science, decreasing the Dark Energy figure of
merit by approximately 40\%. We describe additional metrics focused on key analysis tasks, \cwr{such as}
photometric redshifts and galaxy clustering. \cwr{We propose a new strategy that returns the survey to uniformity at key release years, enabling use of the full survey area and restoring our metrics to the values they would have in a non-rolling cadence without loss of time domain data relative to a rolling survey with the same number of rolling cycles.} 
\ir{This work has informed the third round of optimization of the survey strategy, and the new uniform rolling strategies have been incorporated into the baseline strategy.}
\end{abstract}



\section{Introduction} \label{sec:intro}


The \cwr{NSF-DOE} Vera C.~Rubin Observatory offers a unique product of telescope collecting area and field-of-view, enabling a rapid survey of the sky visible from Cerro Pachón, Chile \citep{2019ApJ...873..111I}.  \cwr{The observatory will carry out the} Legacy Survey of Space and Time (LSST) \cwr{,which will} re-visit observable sky regions at an average cadence of once per three nights, building a ten-year ``movie" of photometric and astrometric variations along with incredibly deep co-added images in the  $ugrizy$ filters.  The LSST science requirements necessitated the construction of a wide-field, high-throughput, fast-slewing telescope, but left \ir{substantial} flexibility in the detailed survey strategy of which filter and sky location to image at each moment in time \cwr{over the ten year survey}. 

Building on many years of work by the LSST scientific community on the impact of observing strategy on the science output of LSST \citep[e.g.,][]{2017arXiv170804058L, scoc_phase1}, Rubin Observatory established a Survey Cadence Optimization Committee (SCOC) to develop a survey strategy that efficiently addresses a wide range of science goals, including the four scientific ``pillars'' of LSST: understanding the nature of dark matter and dark energy, exploring the changing sky, studying the formation and structure of the Milky Way, and cataloging the solar system. The community at large was tasked with defining scientific metrics relevant to their specific science cases and with providing feedback to the SCOC regarding the proposed observing strategies (as well as developing new ones). This years-long process has led to many improvements in the planned survey strategy \citepalias[e.g.,][]{scoc_phase1, scoc_phase2, scoc_phase3}. The survey optimization process is documented in a series of papers covering the development and performance of new metrics on several proposed observing strategy simulations \citep[one example is][while the entire series of community studies are presented in an online repository\footnote{\href{https://survey-strategy.lsst.io/scoc/community.html}{https://survey-strategy.lsst.io/scoc/community.html}}]{2022ApJS..258....1B}. 
One major conclusion from this process was that more frequent re-observations of some parts of the sky (an observing strategy coined ``rolling cadence'') was a highly desirable feature of the LSST since it could significantly improve its science return \cwr{for transient science} \citep{2017arXiv170804058L}.
Early optimization of the cadence strategies focused on determining suitable versions of rolling \citepalias{scoc_phase2}. 
Implementing higher cadence observations for a given part of the sky enables sampling a wider range of timescales for time domain phenomena.

\ir{Given the strong gains for transient science, the SCOC recommended a two-stripe rolling strategy as the baseline \citepalias{scoc_phase2}, which leads to emphasizing half of the observable sky in any given observing season and then flipping the emphasis in the subsequent year, switching back and forth until the end of the survey. The rolling would
start after at least one year of observations without this rolling strategy employed (a ``no-rolling'' cadence) to ensure a stable and homogeneous survey start that supports precise photometric calibration and template creation for transients. Furthermore, to ensure that the part of the observable sky that is de-emphasized is still sampled, but less frequently than the emphasized one, the SCOC recommended a ``rolling \irtwo{weight''} of 0.9. \irtwo{This weight is the relative fraction of images the scheduler is asked to schedule on the active versus inactive parts of the sky; however, since there are many other constraints in the scheduler, the actual fraction of images within the active part of the sky is typically below this value.}
More recent SCOC work addresses other axes of optimization or realism of the survey simulations, such as weather conditions and the exact timing of the rolling observations \citepalias{scoc_phase3}.} 

The LSST Dark Energy Science Collaboration (DESC\footnote{\href{http://www.lsstdesc.org}{http://www.lsstdesc.org}}) is one of eight science collaborations planning large-scale analyses of LSST data.  DESC's primary goal is to \cwr{study fundamental cosmological parameters with LSST, including determining} the nature of the dark energy that is causing the expansion of the universe to accelerate.  The large-scale cosmological analysis planned by DESC requires careful attention to a long list of systematic errors, many of which depend directly on the final choice of survey strategy \citep{DESCSRD}.  DESC \cwr{produced} several papers and white papers that describe these concerns, including \citet{2016ApJ...829...50A}, \citet{2018arXiv181200516S}, \citet{2018arXiv181200515L}, 
and \citet{2022ApJS..259...58L}.  
\ir{This work included optimizations \cwr{for all cosmological probes that DESC plans to use (both static and transient), with the needs of} Type Ia supernovae
making 
the case for a rolling cadence strategy.}
\ir{The initial DESC engagement with the survey cadence optimization process included the optimization of rolling for transient science but did not quantify the importance of uniformity across the LSST footprint.  
Any non-uniformities in imaging depth imposed by the overall survey strategy will create spurious fluctuations in galaxy number counts that mimic large-scale structure.  
The current generation of Stage III dark energy surveys has illustrated the importance of uniform survey depth for cosmological large-scale structure analyses \citep{Heymans_2021, yan2024kidslegacyangulargalaxyclustering, Elvin_Poole_2018, Andrade_Oliveira_2021, Rodr_guez_Monroy_2022, Nicola_2020, miyatake2023hypersuprimecamyear3, Sugiyama_2023}.}

\ir{In this paper, we describe work completed to optimize the \cwr{LSST observing cadence with respect to the survey uniformity} needed for accurate static cosmological analyses at key years (one, four, seven and ten) when such analyses will be conducted.}
This includes new metrics to supplement those used in earlier optimization phases, their application to different versions of rolling cadence, and a new ``uniform rolling'' approach that offers a strategy more suitable for static cosmological analyses.
These developments were key to the feedback provided by DESC to the SCOC in the ``Phase 3 optimization'' of the survey strategy \citepalias{scoc_phase3}, and led to a new set of recommended strategies that leverage uniform rolling and are backed by quantitative metrics for both static and transient cosmological science.

\ir{The rest of this paper is structured as follows:  
\autoref{sec:baseline} introduces the survey strategy optimization procedure and some default metrics used to perform such an optimization. We also give a  
pedagogical description of the Phase 2 \cwr{(\vtpf)} survey strategy and the uniform rolling approach that informed the development of the Phase 3 recommendations.
  \cwr{In \autoref{sec:metrics}, we describe} our approach to quantifying how the resulting galaxy number density and photometric redshift quality depend upon the imaging depth, and introduces new metrics developed to measure the impacts of survey non-uniformity upon the science achievable with LSST.  
\cwr{In \autoref{sec:disc/conc}, we  discuss} our findings \cwr{and conclusions}.  } 



\section{Survey Strategy Optimization for the Rubin Observatory}
\label{sec:baseline}

This section presents the key concepts of the LSST survey strategy optimization, and the specific survey simulations considered to study \cwr{survey uniformity} for static cosmological analyses. \ir{The strategies described in this section and used for the analysis in subsequent sections are also summarized briefly in \autoref{tab:strategies}}.

\ir{\subsection{Rubin Operations Simulator}}
\label{sec:opsim}

\ir{LSST survey strategies are simulated by the Operations Simulator}\footnote{\url{https://rubin-sim.lsst.io/}} \ir{(OpSim) of the Rubin Observatory. }

\ir{OpSim simulates the scheduling of observations given various constraints and inputs, such as telescope movements, observing date, and weather conditions. The output is a long sequence of \cwr{telescope pointings with a complete description of the observing conditions and image characteristics} across the LSST survey footprint over the 10-year observation period. These can then be processed by the Metrics Analysis Framework (MAF)\footnote{\href{https://www.lsst.org/scientists/simulations/maf}{https://www.lsst.org/scientists/simulations/maf}} to \cwr{calculate various summary statistics and derived metrics describing the survey and its scientific performance}.
These metrics can be used to assess the performance of the observing strategy. 
For example, MAF is able to create maps of survey conditions and in turn derived metrics (e.g., \cwr{coadded} $5\sigma$ depth) sampled on the sphere \ir{following the Hierarchical Equal Area isoLatitude Pixelisation \citep[HEALPix,][]{healpix2005}} format. This can then be further compressed into scalar summary metrics, for instance, the standard deviation of the depth in a given band over a particular region of the sky.
In what follows, we choose a resolution of $N_{\rm side}=64$ for all sky maps. This is sufficient to capture the smoothly-varying observing conditions on the sky and the physical scales considered in the metrics we developed.
\cwr{In \autoref{fig:rizexptime} we show} an example of a summary statistic (the total $r+i+z$ exposure time) calculated with MAF, displayed as mollweide projection.}

\ir{OpSim \cwr{can simulate observations in any sky area} and observing conditions accessible by Rubin. \cwr{More specifically, OpSim simulates the properties of Rubin images, not the images themselves. Mock images or catalogs of objects can be created by further processing OpSim outputs, as shown in \autoref{sec:simulation}, for example.}
However, cosmological analyses will only make use of a subset of this data, known as the Wide-Fast-Deep (WFD) region, covering $\sim$20,000 square degrees.
In the context of this work, it is defined as the large region left when excluding the deep drilling fields (DDFs) and the mini survey areas. Furthermore, a\cwr{n extinction} cut $E(B-V) < 0.2$ is applied in order to avoid uncertainties from the dust map and the dust extinction law. We also \cwr{use scheduler notes accessible via MAF to} filter out observations at twilight in near-sun positions at high airmass, which are part of the search for solar system objects \citep{seaman2018nearsunsolartwilightsurvey}.} 

\subsection{\cwr{The Phase 2 (\vtpf) survey strategy}}

\ir{Community-designed metrics have been used to evaluate a series of increasingly realistic simulations of the ten-year LSST, including the main WFD survey, pencil-beam DDFs, and additional mini-surveys covering the North Ecliptic Spur, Southern Galactic Cap, Galactic Center, and Target of Opportunity observations.  The result of this process \cwr{was development of} a \cwr{Phase 2 baseline (\vtpf)} survey strategy that meets the primary science goals \cwr{for Y1 and Y10} and will enable a wide range of scientific investigations by the community, as detailed in \citet{2022ApJS..258....1B} and the reports from the SCOC \citepalias{scoc_phase1, scoc_phase2, scoc_phase3}.}

The most prominent features in the \vtpf\ survey strategy for the WFD component
of the Rubin Observatory LSST are the striped bands of varying depth that appear as the survey is executed \citep{yoachim2024}.
These bands are due to the ``rolling'' feature of the survey, which is defined in more detail in \autoref{subsec:rolling}, and which was prioritized in recent versions of the observing strategy\footnote{\url{https://pstn-055.lsst.io/} or \citetalias{scoc_phase2}}. Simply put, rolling is a technique where
different parts of the survey are observed more frequently than others in a way that alternates the low- and
high-cadence areas over time. The net effect of this change is to produce better sampled light curves for time-domain
science at the expense of a less homogeneous survey at intermediate times \citep{2022ApJS..259...58L}. By alternating the areas of low- and high-cadence, one
can eventually restore approximate survey homogeneity. However, the timing of this restoration had not been extensively studied or optimized. This is particularly relevant for static cosmological analyses using data from data releases that suffer from the non-uniformity introduced by rolling, as described in \autoref{subsec:needforuniformity}. 
Thus, strategies that restore uniformity at key intermediate years before the end of the survey would be desirable.

\subsection{The need for survey uniformity}\label{subsec:needforuniformity}

\ir{Stage III dark energy surveys have illustrated the importance of uniform survey depth for cosmological large-scale structure analyses \citep{Heymans_2021, yan2024kidslegacyangulargalaxyclustering, Elvin_Poole_2018, Andrade_Oliveira_2021,Rodr_guez_Monroy_2022, Nicola_2020, miyatake2023hypersuprimecamyear3, Sugiyama_2023}, which typically involve measuring subtle correlations of galaxy overdensities and/or shapes.  Such correlations, including those caused by a non-uniform survey coverage or depth, can also be spuriously induced by fluctuations in galaxy number counts and derived quantities.} 
\ir{Current techniques to handle this include regression or deprojection against templates of potential systematics and forward modeling \citep[e.g.,][]{2024arXiv240112293B}.
Typically, systematics mitigation is only exact when the contamination to the density or shear fields is a linear function of observational templates across the survey footprint. 
Some techniques are effective at capturing non-linear contamination, but this gets more challenging as the contamination increases, and it is a function of the galaxy samples considered.
Residual systematics can propagate and jeopardize the subsequent cosmological analysis.
Even though uniformity cannot be achieved perfectly because of unavoidable contributions from weather fluctuations, foregrounds, and other observational or instrumental systematics, these contributions come on top of any obvious large-scale patterns, such as the stripes created by rolling cadence.
Thus, it is critical to avoid such large patterns if possible.}

A known consequence of rolling cadences is that uniformity in survey depth is only achieved after a cycle of alternating emphasis completeness.  In the case of the adopted baseline LSST strategy \cwr{in early 2024, \vtpf,} rolling cycles were interwoven in order to optimize time domain science, meaning that the only uniform LSST data releases would be after the first year (before rolling starts, \cwr{with rolling beginning anywhere from 12-18 months into the survey for the range of strategies we analyze}) or at the end of the ten-year survey (if an integer number of rolling cycles reach completion). 
In addition to the advantages of achieving uniformity at Year 10, it is desirable to obtain near-uniformity to enable full cosmological analyses at intermediate times, e.g., Years 4 and 7.  The advantages of this approach include building up expertise and infrastructure to work with a new and complicated dataset, enabling discoveries with early data, training and recognizing the work of early-career scientists, and continuity in the expertise capitalized in the Science Collaborations. This points to a timescale of $\sim$3 years for an end-to-end cosmological analysis, as demonstrated by previous surveys, and 
motivates DESC's intent to perform cosmological analysis on \cwr{uniform} LSST data releases at the end of Years 1, 4, 7, and 10.  As noted above, the \cwr{Phase 2 baseline (\vtpf)} observing strategy did not achieve uniformity at such intermediate times.  
\ir{However, we can resolve this issue by altering the way rolling is implemented as described below.}

\subsection{Rolling cadence}\label{subsec:rolling}

There are several rules-of-thumb and key facts to keep in mind when thinking about rolling surveys as implemented
for the LSST WFD by the survey scheduler\footnote{\url{https://github.com/lsst/rubin_scheduler}}.

\begin{enumerate}
    \item The survey scheduling software defines a ``season'' as a function of time and location for all parts of the sky.
    For the purpose of this work, a location's season increments to a new whole number three months before the sun passes through the right ascension of the location. \cwr{Note that there are other possible definitions of a season, differing by a phase shift, but this definition is used internally by the scheduler and so is most useful here.} The scheduler
    is instructed to observe a position starting from six months into its season and ending at twelve months in its season. \cwr{(Note that depending on the declination of the location, it may become visible earlier than six months into its season or into the next season at twelve months, but the scheduler does not currently account for this.)}
    \item When the survey is rolling, a given location of the sky is observed at higher or lower cadence than usual for its
    entire season, as opposed to say a full calendar year in time. This choice increases the length of high-cadence light curves
    for transient objects. It also has the net effect that it takes $\sim$2.5 years for the full LSST WFD to execute a single rolling
    cycle (i.e., every location on the sky has both a high- and a low-cadence season.). The extra six months comes from ``partial''
    seasons at the start (or end, depending on \cwr{one's} point of view) of the nominal two years.
    \item High- and low-cadence areas have to be balanced across the survey in order to maintain the total rate of observations. Very roughly,
    it is most effective to balance them in declination with the low- and high-cadence regions having the same total area and stacked
    on top of one another. One of the regions is observed at high cadence and the other at low cadence for a full season. Then \cwr{i}n the next season,
    the low- and high-cadence areas are swapped to bring the total survey back to a uniform depth.
\end{enumerate}

\begin{figure*}
    \centering
\includegraphics[width=\textwidth]{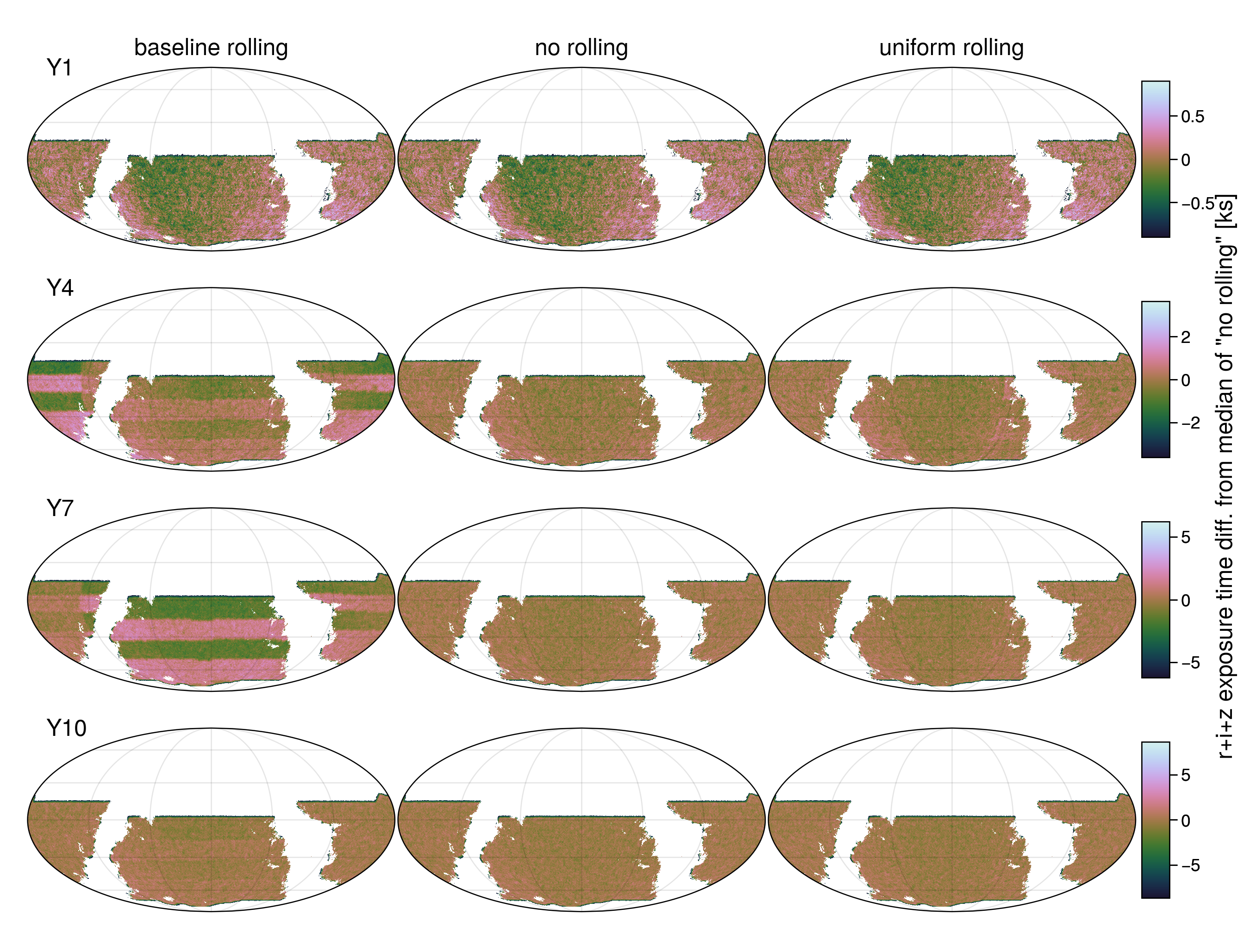}
    \caption{
      \ir{Difference in $r+i+z$ exposure time from median of the \noroll\ survey in kiloseconds. Edges of the WFD survey with total exposure time 
      less than an average of 700 seconds per year\cwr{, arising due to the incomplete coverage at the edge of a survey with dithers between pointings,} have been cut from the plot to improve clarity. The left column shows the \vtpf\
      survey, the middle column shows the \noroll\ survey, and the right column shows the \vtpfur\ survey. The
      rows show LSST at years one, four, seven, and ten. Unlike the \cwr{Phase 2 baseline (\vtpf)} survey where rolling stripes are still apparent, the
      uniform rolling survey returns to \cwr{similar performance as} the \noroll\ survey at years four, seven, and ten. Year one is similarly
      uniform in all three surveys by design\cwr{, since rolling does not start until after that point}. }
    }
    \label{fig:rizexptime}
\end{figure*}
Beyond these rules-of-thumb, the low- and high-cadence time periods can be arranged in any way one would like. 

\vspace{0.2in}
\ir{\subsection{Uniform rolling}}
\label{sec:uniformrolling}
As outlined in \autoref{sec:intro}, given the evidence that substantially non-homogeneous coadds can cause problems that are challenging to mitigate during the large-scale structure analyses, it is \cwr{imperative} to explore variations that retain the benefits of rolling for time-domain science, while returning as closely as possible to a homogeneous survey at key data releases to be used for static science.

While \cwr{observing cadences that do not exhibit rolling (so-called `no-rolling' strategies)} not under consideration for LSST, it serves as a useful data point illustrating the most homogeneous survey possible with the given fixed exposure times and weather/downtime model (i.e., so that we do not compare with an unrealizable case of a perfectly homogeneous survey).  


The core idea behind uniform rolling is that for each location in the survey, we insert a full season of \cwr{either a uniform or a `no-rolling' survey observations in between the rolling cycles}. Then, by adjusting when we start rolling and during which season a given location executes a given cadence, we can arrange the survey to return to the \cwr{observing strategy that does not have rolling} at specific times. This procedure is not completely flexible in that we cannot arrange the survey to be uniform at arbitrary times, and it takes three years to execute a full uniform rolling cycle. \cwr{We compare these strategies in \autoref{fig:rizexptime} of \autoref{subsec:strategies} and describe the procedure for generating the uniform rolling in \autoref{app:uroll}.}



While the subject of this work is the implications of rolling cadence for survey uniformity, we note that the Phase 3 report of the SCOC \citepalias{scoc_phase3} commented that based on feedback at the time, adoption of uniform rolling (instead of the previous rolling cadence) had ``limited detrimental impact on time-domain probes'', and encouraged further scientific exploration by the time-domain community to further characterize these strategies.  This endorsement also helps  motivate the further justification and exploration of uniform rolling strategies for static science.

\subsection{Simulated observing strategies considered}
\label{subsec:strategies}
\ir{We focus on four specific simulations of the LSST cadence with different prescriptions for rolling: the \vtpf\ simulation that implemented a version of rolling cadence that (we will show) leads to non-uniformity across the sky at intermediate years; the \vtpfur, which is an implementation of the v3.4 cadence with a rolling cadence procedure that yields a uniform survey at years 4 and 7; the \vfpz\ strategy which includes the uniform rolling procedure and \vfpzfc, a version of the v4.0 survey that includes four cycles of rolling (compared to the baseline of three cycles) to show the dependence of the survey uniformity on the number and duration of rolling cycles. } \cwr{All of these strategies are based on two-stripe rolling cadence since it is the preferred type of rolling since the SCOC Phase 2 recommendations \citepalias{scoc_phase2} and is the only type of rolling considered at present.}
\begin{deluxetable}{lp{10cm}}
\tablecaption{The OpSim strategies considered in this work.}
\label{tab:strategies}
\tablehead{
  \colhead{Name} & \colhead{Description}
}
\startdata
\vtpf & The SCOC Phase 2 baseline simulation (from December 2023) including half-sky (2-\cwr{stripe}) 0.9-weight rolling cadence \cwr{in the WFD survey}.\\
\vtpfur & A modification of the Phase 2 baseline that prioritizes uniformity at years 4, 7 and 10. \\
\vfpz & The SCOC Phase 3 baseline simulation (from September 2024) including 3-cycle uniform rolling cadence on the WFD.\\
\vfpzfc & A modification of the Phase 3 baseline implementing 4-cycle rolling.\\
\hline
\noroll & A version of the Phase 2 baseline that does not include rolling, used as a comparison to estimate the impact of ignoring rolling.\\
\hline
\enddata
\end{deluxetable}
\ir{We note that there are a variety of differences between v3.4 and v4.0\footnote{Described in full at \url{https://community.lsst.org/t/scoc-phase-3-recommendation/9429}}, such as the change of mirror coatings, the inclusion of Target of Opportunity programs, longer $u$-band exposures, better estimates of downtime in Year 1, and more pessimistic weather patterns.
These general differences do not relate to the rolling cadence, and outside of rolling they only have minor effects on uniformity, so we will not discuss them in detail here.
In practice, it is best to compare strategies of a given version with each other and avoid inter-version comparisons.
However, the comparison of two v3.4 and v4.0 simulations helps to assess the effect of these more recent changes on the metrics developed.
The four strategies chosen are sufficiently representative of the range of simulations of Rubin's observing procedure and highlight the power of our metrics to discern between fine-grained choices in observing strategy.

Finally, in order to illustrate the effect of rolling itself in all years, some of our comparisons will include a fifth simulation: one with no rolling as described in \autoref{subsec:rolling} (from the suite of v3.4 runs\cwr{: \noroll}).  This is a technical comparison only,  as a non-rolling strategy is not currently under consideration for LSST.} 

\cwr{In \autoref{fig:rizexptime}, we illustrate the various arrangements used in this work, comparing them to \cwr{the} \noroll\ \cwr{strategy} (shown in the middle column) where rolling is turned off completely. 
First, the left-most column of this figure shows the \vtpf\ survey's differences in total exposure time from the median of the \noroll\ survey in kiloseconds. We use the combined exposure time in the $r$, $i$ and $z$ bands as our measure of depth. \ir{This is because these bands are the most relevant for cosmic shear measurements \citep[e.g.,][]{2023OJAp....6E..17S}.} The \cwr{\vtpf} survey executes four rolling cycles in time and uses a two-stripe pattern across the sky, alternating high- and low-cadence seasons in these stripes. These rolling sequences start after a full $\sim1.5$ years of operation and are very apparent in the survey depth at all years past Year 1 (rows two through four). \ir{We only show Years 1, 4, 7 and 10, which are years during which \irtwo{DESC} cosmological analysis are planned and thus \cwr{for which} uniformity should be maximized.}}

\cwr{Second, the right-most column of \autoref{fig:rizexptime} shows the \vtpfur\ strategy proposed in this work. This rolling
strategy is designed explicitly to bring the survey back to the \noroll\ survey after each rolling cycle. In order to achieve this increased level of uniformity, the uniform rolling strategy can only execute three rolling cycles, as opposed to four in the baseline. Further, this strategy can
only match the \noroll\ survey every three calendar years. The uniform rolling strategy examined in this work has been set to match the \noroll\ survey at Years 4, 7 and 10. In between these years, this survey also shows rolling stripes in the depth across the survey.}

\section{Uniformity metrics and Observables} 
\label{sec:metrics}

In this section, we describe the new metrics introduced to study \cwr{survey uniformity} from the standpoint of static cosmological analyses. \ir{Doing this requires} mock galaxy catalogs that allow us to model fluctuations of galaxy number densities and mean redshift in tomographic redshift bins as a function of six-band depth. 
We first describe the mock galaxy photometry catalogs, which are used as the basis for some of the metrics introduced in the rest of the section in \autoref{sec:simulation}. From the mock catalogs, we quantify the sensitivity of the number density and the mean redshifts of galaxies in six tomographic bins defined by their photometric redshifts in \autoref{sec:derivatives}. We then describe in detail \irtwo{four} different metrics to quantify the effects of uniformity: Stripiness (\autoref{sec:stripinessmetric}), Area at risk (\autoref{sec:fommetric}), Tomographic $\sigma_8$ bias (\autoref{sec:sigma8}) and the Mean $z$ metric (\autoref{sec:meanz}).

\subsection{Mock galaxy catalogs}
\label{sec:simulation}
\input{simulation}

\subsection{Quantifying sensitivity of density and redshift to depth}
\label{sec:derivatives}
\input{derivatives}

\subsection{Metric: Stripiness (\texttt{StripinessMetric})}\label{sec:stripinessmetric}


\ir{Galaxy detections and shape measurements are affected by depth in the six Rubin bands.
These get mixed, possibly non-linearly, by any downstream multi-band algorithms, for example when determining photometric redshifts.
Thus, galaxy clustering and weak lensing will non-trivially depend on the six-band depth fluctuations.  
However, cumulative $r+i+z$ exposure time is the lead contributor to the signal-to-noise in galaxy clustering and weak lensing because these three bands are the bands with the highest signal-to-noise and the most relevant for the detection of galaxies and the measurement of shapes and colors \citep[e.g.,][]{2023OJAp....6E..17S}.
Thus, it can be used as a summary statistic to study observing strategies.}

\begin{figure*}
    \centering
    \includegraphics[width=0.9\textwidth]{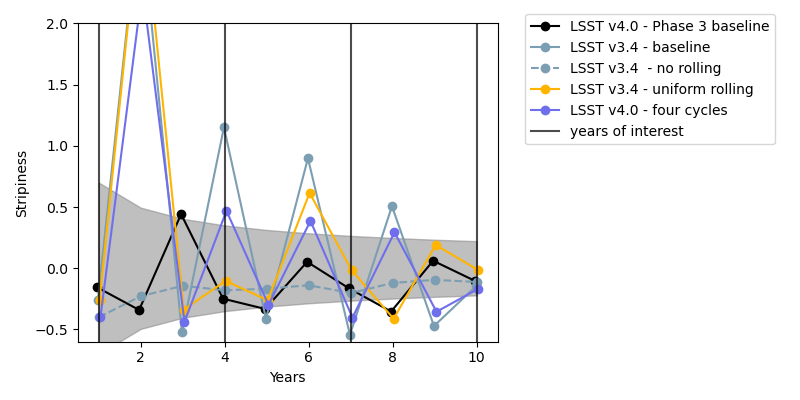}
    \caption{
      Quantitative assessment of the amount of exposure time variation vs.\ year in different observing strategies, as measured \cwr{using} the stripiness metric of \autoref{sec:stripinessmetric}.  The test statistic plotted on the vertical axis effectively measures the fractional difference between the fractional variations in depth in the northern and southern Galactic regions, with a value of 0 indicating that the two are the same, as expected for a perfectly uniform survey.  The \ir{grey} shaded envelope indicates the region for which we consider the stripe features to be negligible (meaning \ir{the bias can be modeled and controlled using current analysis algorithms). The dashed grey line for the \noroll\ case can be used to estimate the approximate magnitude of depth fluctuations that might originate purely from random noise.} }
    \label{fig:stripiness}
\end{figure*}

The striped patterns in the cumulative exposure time due to rolling, as shown in \autoref{fig:rizexptime}, are visually obvious.  However, we need a clear and quantitative metric to find them in an automated way.  To define such a metric, we \cwr{observe} that these patterns result in different distributions of cumulative exposure time in the northern and southern galactic regions.  This observation \cwr{leads} to definition of a `stripiness' metric that carries out the following steps:
\begin{enumerate}
    \item We define a `fractional scatter' in the cumulative  distribution of $r + i + z$ exposure times within the northern and southern galactic regions, using the ratio of the median absolute deviation (MAD) to the median. 
    \item \ir{We take the fractional difference between those fractional scatters, and check whether it is smaller than $0.7/\sqrt{\text{year}}$, a threshold that was empirically demonstrated to correspond to the clear presence of striping features in the detected galaxy density and that exceeds the north-south differences that can arise due to noise fluctuations by a factor of 3. }
    \irtwo{This fractional difference is shown in \autoref{fig:stripiness}.  The $0.7/\sqrt{\text{year}}$ threshold is indicated with a grey shaded envelope.}
    \begin{itemize}
    \item If yes, meaning the northern and southern regions have consistent distributions of cumulative exposure time (within \irtwo{3$\sigma$}), we consider there to be no significant evidence for stripiness. \irtwo{This is the case for all points falling within the grey shaded region \cwr{i}n \autoref{fig:stripiness}.}
    \item \ir{If no, meaning the differences in northern and southern regions exceeds by a factor of 3 what is expected due to noise (as assessed based on year-to-year fluctuations for a strategy without rolling), we consider this as significant evidence for residual stripiness.} \irtwo{This is the case for all points falling outside the grey shaded region \cwr{i}n \autoref{fig:stripiness}.}
    \end{itemize}
\end{enumerate}

 We show the value of this test statistic for several strategies as a function of year in \autoref{fig:stripiness}. As highlighted \cwr{i}n the plot, for the years that are especially of interest for static science analysis (1, 4, 7, 10), strategies with no rolling -- which are not a serious candidate at this stage -- or with uniform rolling generally have values of the stripiness metric that would be classified as near uniform, while other strategies do not consistently achieve this benchmark.  

\subsection{Metric: Area at risk (\texttt{UniformAreaFoMFractionMetric})}\label{sec:fommetric}

The `area at risk' metric is devised based on the fact that past cosmological weak lensing and large-scale structure analyses have required substantial effort to remove systematic large-scale power induced by observing condition-related issues \citep[e.g.,][]{2022MNRAS.511.2665R,2024arXiv240112293B}.  \cwr{For example, in the very recent Dark Energy Survey Year 6 analysis \citep{2025arXiv250907943R},  even with the very homogeneous (in depth) DES dataset, the authors had to remove portions of the footprint that are outliers in certain survey properties to stabilize the calibration, essentially making an area versus systematics decision.  While they only lost a few percent of the area, the area cuts drastically increase in severity with modest increases in needs for systematics control.  This is therefore} a pressing issue for LSST given that its higher statistical constraining power implies a need for \cwr{substantially} tighter control of systematic uncertainties than for precursor surveys.  This is in practice an issue on all spatial scales, but especially on large scales where the cosmological clustering amplitude is low.  Given the challenge in removing these impacts even when the observing strategy gives a constant exposure time across the survey, it is not safe to assume that we will definitely have systematics mitigation strategies that can also remove the large-scale power induced in the coadds by rolling.  
We therefore assume that for coadds that have the characteristic stripe features of rolling, we may need to divide the survey into subregions with different characteristic depths, and use the subregion with the largest cosmological constraining power for the analysis. \ir{This is a conservative assumption, as a proper analysis accounting for the different properties in the interleaved regions of substantially different depth would be very complex to develop and also computationally demanding.}

\begin{figure}
    \centering
    \includegraphics[width=0.49\textwidth]{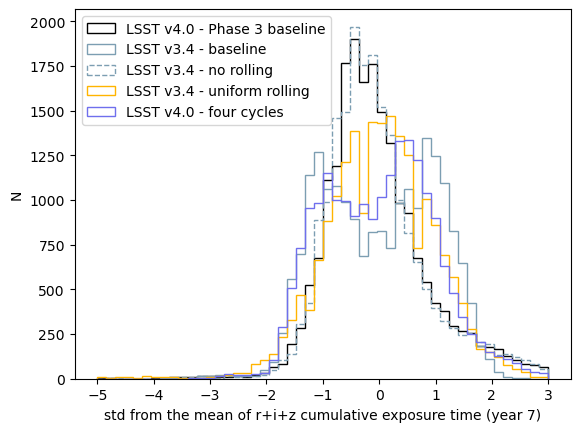}
    \includegraphics[width=0.49\textwidth]{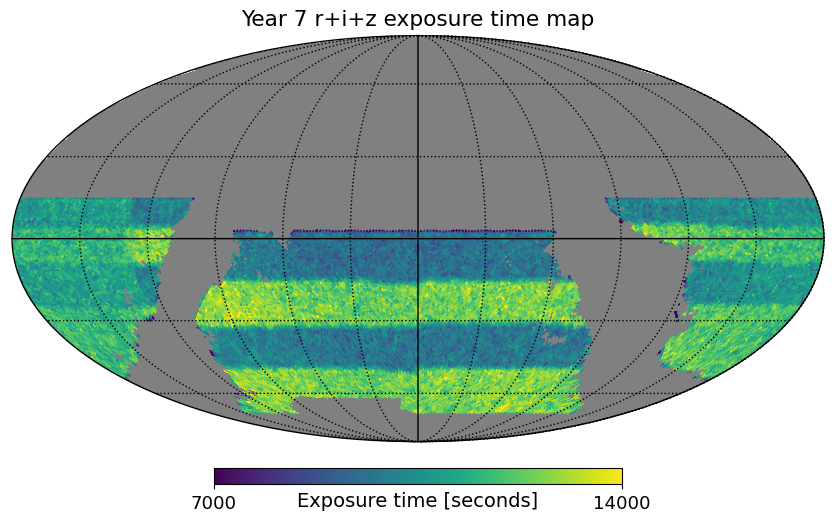}
    \includegraphics[width=0.49\textwidth]{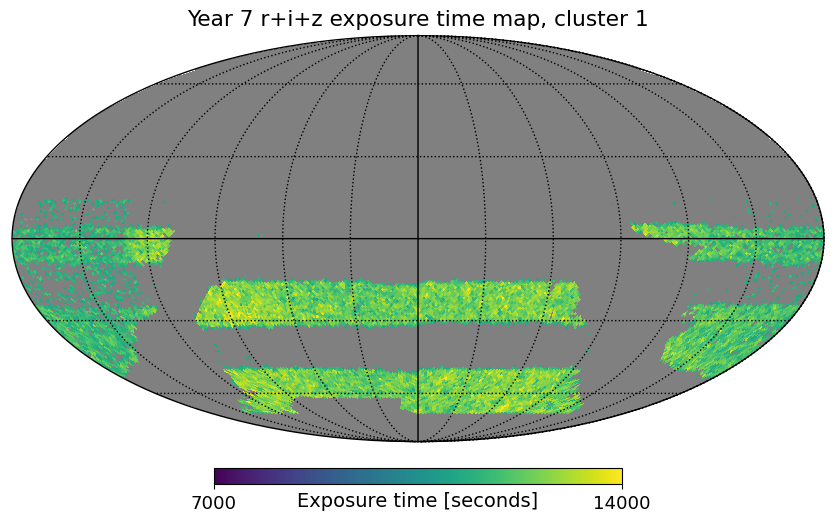}
    \includegraphics[width=0.49\textwidth]{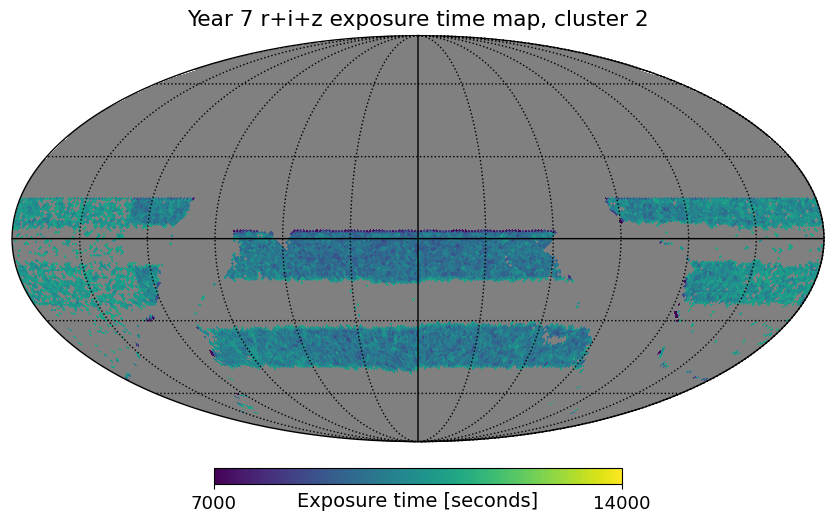}
    \caption{Each panel shows a different input to the area at risk metric.  {\em Top left:} The histogram of $r+i+z$ cumulative exposure time values (from a HEALPix map with $N_\mathrm{side}=64$) after 7 years of the survey have passed.  The exposure time values are in units of standard deviations from the mean across the survey.  The double-peaked histogram for the \vtpf\ strategy is caused by the `striping' effect induced by rolling.  As shown, the uniform rolling approach implemented in \vfpz\ produces a single narrow peak, similar to what is achieved without any rolling.  {\em Top right:} A visualization of the exposure time map used in the top left panel for the \vtpf\ strategy.  {\em Bottom left, right:} These show the clusters identified by the area at risk metric.  As shown, much of the survey divides cleanly -- however, in a few places it is shredded by the segmenting algorithm.  While we would not use such a discontinuous region for large-scale structure analysis, the shredding has a small impact on the estimated areas for the deeper and shallower regions, which is the \cwr{quantity of focus here}.}
    \label{fig:areaatrisk-diagnostics}
\end{figure}

This logic motivates a simple definition for this metric, based around the cosmological constraining power for weak lensing and large-scale structure analyses.  We begin with the existing 3$\times$2-point Figure of Merit (FoM) metric already in MAF, \texttt{StaticProbesFoMEmulatorMetric}, which quantifies the dark energy constraining power for a joint weak lensing and large-scale structure analysis (`$3\times 2$-point analysis'), as a proxy for its overall cosmological constraining power \citep{2022ApJS..259...58L}.  The steps in computing our new `area at risk' metric for a particular strategy in a particular year are as follows:
\begin{enumerate}
    \item We apply cuts to get the extragalactic WFD area as in \autoref{sec:opsim}, and get the combined $r+i+z$ exposure time \texttt{RIZDetectionCoaddExposureTime} as a function of position on the sky.  Th\cwr{e}se bands will likely dominate the weak lensing shear inference \citep{2023OJAp....6E..17S} due to their signal-to-noise ratio for typical weak lensing galaxies and the prospects for well-controlled point-spread function modelling.
    \item We analyze the distribution of $r+i+z$ exposure time to identify whether there are residual rolling features in them (see top left panel of \autoref{fig:areaatrisk-diagnostics} for an example of the impact of rolling features on the exposure time distribution).  In practice, we use the stripiness metric from \autoref{fig:stripiness} to make a binary decision as to whether stripes are significant or not. 
    \begin{itemize}
    \item If no significant stripe features are identified, then the assumption is that we can use the full extragalactic area for cosmological weak lensing, and the metric returns 1, meaning that the constraining power is unaffected.
    \item If stripes are identified, the metric uses a $k$-means clustering algorithm from \texttt{scikit-learn} to segment the area into shallower and deeper regions each with narrower distributions of exposure time (see the bottom row of \autoref{fig:areaatrisk-diagnostics} for an illustration for a single strategy and year).
    We identify which of the regions gives the highest cosmological constraining power via the $3\times 2$-point FoM, which considers the median depth and total area of each region.  We then assume that that is the region we will use for cosmological analysis, and the metric returns the ratio of its FoM to the FoM we would have gotten if we could have used the entire area, effectively treating the rest of the area as having been lost due to unmitigated systematics. \cwr{The reason we assume the rest of the area is lost, rather than that we do separate analyses for each area and then combine the results at the end, is because the two interleaved areas have a non-negligible covariance (both in terms of the real clustering and LSS, and in terms of systematics) and we have no solid basis in the literature for how to account for that covariance properly.}
    \end{itemize}
\end{enumerate}

\begin{figure}
    \centering
    \includegraphics[width=0.9\textwidth]{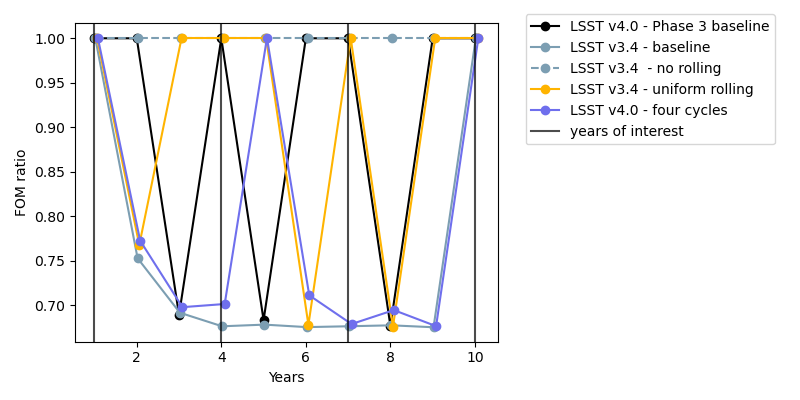}
    \caption{This plot shows the value returned by \texttt{UniformAreaFoMFractionMetric}, the metric presented in \autoref{sec:fommetric}.  For strategies and years in which the coadds lack evidence for rolling stripe features in the exposure time maps (summing $r$, $i$, and $z$ exposure times), the metric returns 1, meaning that no area has to be cut to achieve uniformity and the cosmological constraining power is unaffected.  For years with rolling stripe features, the metric segments the area into regions with greater and lesser depth, and returns the ratio of the $3\times 2$-point FoM for the region with more constraining power divided by what we would get using the entire area.   As shown, for the \vtpf\ strategy, there is evidence for rolling stripes in all years from 2-9, resulting in considerable loss of cosmological constraining power.  However, the uniform rolling strategies \cwr{(\vtpfur\ and
  \vfpz)} by design have uniform coadds at years 4, 7, and 10 (highlighted with vertical lines), and this results in the metric value returning to 1 at those years.}
    \label{fig:areaatrisk}
\end{figure}

As a reminder, we only plan cosmological weak lensing analyses at particular data releases (for the purpose of this discussion, years 1, 4, 7, and 10).  \cwr{In \autoref{fig:areaatrisk}, we 
 show} the performance of various 
 simulations using this metric, with vertical lines highlighting the years of interest. As shown, for the \vtpf\ strategy, there is evidence for rolling stripes in all years from 2--9, resulting in considerable loss of cosmological constraining power.  In contrast, the \noroll\ strategy results in a metric value of 1 in all years.
 Finally, the uniform rolling strategies (\vfpz, \vtpfur) by design have uniform coadds at years 4, 7, and 10 (highlighted with vertical lines), and this results in the metric value returning to 1 at those years.

\subsection{Metric: tomographic $\sigma_8$ bias (\texttt{TomographicClusteringSigma8biasMetric})} 
\label{sec:sigma8}

\cwr{In \autoref{sec:fommetric} we}  presented a metric based on a conservative assumption that with sufficient non-uniformity we might exclude part of the area from a cosmological lensing and clustering analysis.  This section, in contrast, is based on an optimistic view that we will use the entire extragalactic area and will successfully model the impacts of non-uniformity at the 90\% level.  Reality is likely to be between these two scenarios.

In order to quantify the effect of the observing strategy on the inferred cosmological parameters, one would ideally need realistic simulations including cosmology, astrophysics (e.g., galaxy populations), and photometric noise \citep[e.g.,][]{2024MNRAS.535.2970H} -- or even image simulations. 
However, in galaxy clustering, most forms of contamination (e.g., spatially varying depth, stars, blending) will manifest as extra power in the measured angular power spectra.
Thus, the main parameter affected  will be $\sigma_8$,  the amplitude of the linear matter power spectrum on the scale of $8 h^{-1}$Mpc, which is almost exactly a proportionality factor in the projected angular power spectra of tomographic galaxy density maps.
We develop a metric making several simplifying assumptions in order to directly convert maps of depth fluctuations into an analytic estimate of the bias made on $\sigma_8$, bypassing the need for expensive simulations, and allowing us to rapidly compare observing strategies. 
\ir{For a more in-depth study of large-scale structure systematics and their effect on cosmological parameters, see \cite{awan2024impactlargescalestructuresystematics}.}

\begin{figure}
    \centering
    \includegraphics[width=0.98\textwidth]{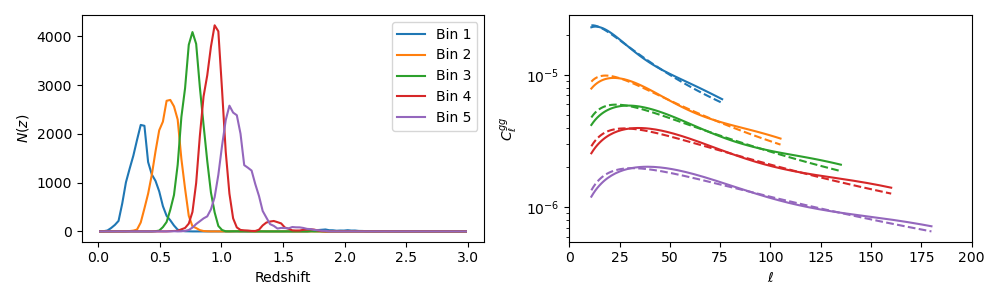}
    \caption{Redshift distributions and fiducial angular power spectra calculated for the five tomographic redshift bins considered. The dashed lines show polynomial fits of $\log C_\ell^{gg}$ as a function of $\log \ell$, which are easier to manipulate than the tabulated angular power spectra, and sufficiently accurate for our purposes. Further details are in \autoref{sec:derivatives} and \autoref{sec:sigma8}.}.
    \label{fig:sig8_nz}
\end{figure}

First, we employ CCL \citep{2019ApJS..242....2C} with a fiducial cosmology $(\Omega_c=0.25, \Omega_b=0.05, h=0.7, n_s=0.95, \sigma_8=0.8)$
to calculate angular power spectra using the Limber approximation \citep{10.1093/mnras/stab3578}.  We do this for 5 of the 6 tomographic redshift bins\cwr{, dropping the first redshift bin since it is not part of the fiducial analysis \citep{DESCSRD}, as}  presented in \autoref{sec:derivatives}. 
Since the signals computed use the nonlinear power spectrum, the signals are not exactly proportional to $\sigma_8^2$ on all scales, but they are close. 
We assume a linear galaxy bias of unity\cwr{.  While galaxy bias is often degenerate with $\sigma_8$, we ignore this issue here, especially since other observables (e.g., cross-correlations with CMB lensing) can provide a direct handle on galaxy bias and lift this degeneracy.}
We focus on large scales by setting a maximum harmonic multipole of $\ell_\mathrm{max}=k_\mathrm{max}\chi(z)$  with $k_\mathrm{max} = 0.1h^{-1}$Mpc to consider scales which can be modeled with current theoretical frameworks \citep[e.g.,][]{2024JCAP...02..015N}.
This also guarantees that the Poisson noise is subdominant.
We set $\ell_{\min}=10$ to avoid the largest scales, which will likely be the most contaminated and thus the most challenging to employ.
The redshift distributions (taken from the mocks presented in \autoref{sec:simulation}) and fiducial angular power spectra of these five redshift bins are shown in \autoref{fig:sig8_nz}. 
We further assume \cwr{a} Gaussian covariance, which neglects the coupling between $\ell$ modes due to the survey mask\footnote{\ir{While this coupling can redistribute power across scales \citep[see, e.g.,][]{Leistedt_2013}, this effect is only severe on the largest scales. Because we cut the $\ell<10$ modes, the redistribution of power will only cause negligible shifts in the final metric. And since we use the same mask in all cases considered, the relative differences between observing strategies will be unchanged.}}.
As a result the total variances measured in the tomographic overdensity maps are Gaussian distributed and independent, as described in \autoref{appsigma8}.  These cosmological quantities are the initial starting point for the metric, which then does the following to estimate the contamination for a given observing strategy and year:
\begin{enumerate}
    \item We calculate the density derivatives described in~\autoref{sec:derivatives}: when applied to each pixel of the maps of photometric depth in the six bands presented in~\autoref{sec:opsim}, we construct maps of spurious density fluctuations in each of the five redshift bins (HEALPix maps at the same resolution).
    \item We calculate angular power spectra \cwr{using Healpy} and convert them to a total spurious variance, as described in \autoref{appsigma8}.
    \item We assume that 10\% of this total spurious variance will contribute to the final bias on $\sigma_8$, because systematics mitigation techniques (e.g., taking advantage of templates, see \citealt{2021MNRAS.503.5061W}) are able to remove the majority of the contamination, \ir{but there is typically some evidence for small residuals \citep[e.g., figure~14 in][]{10.1093/mnras/staa1621}}. \ir{The 10\% number is meant as a heuristic to reflect this situation rather than as a precise estimate, since it depends in detail on the methods used and systematics that are present.} In reality, the correcting power would be a function of scale and redshift, but this is difficult to predict without realistic simulations.
    \item We convert this into a posterior distribution on $\sigma_8^2$.
The latter is Gaussian due to the approximately multiplicative nature of this parameter and the Gaussian likelihood function, as detailed in \autoref{appsigma8}.
\item We calculate a bias (relative to the statistical uncertainty) on $\sigma_8$ as a final output value for the metric.
\end{enumerate}
%

\begin{figure}
    \centering
    \includegraphics[width=0.65\textwidth, trim={0cm 0 12.5cm 0},clip]{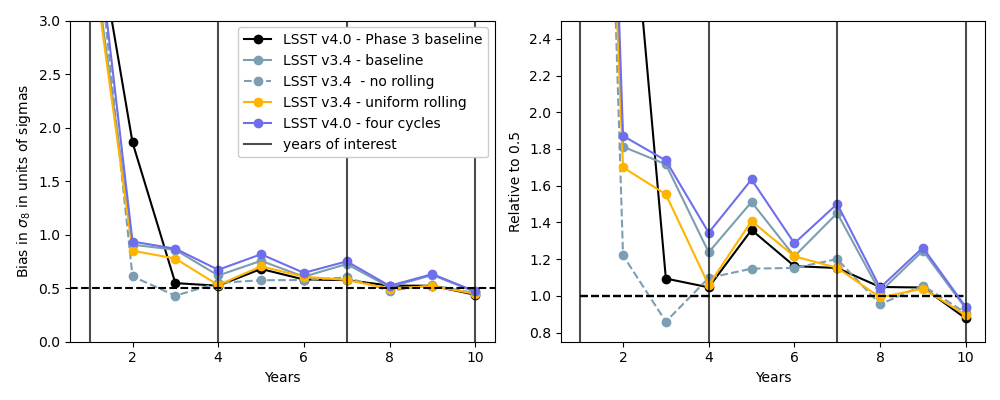}
    \caption{This plot shows the metric value returned by \texttt{TomographicClusteringSigma8biasMetric} defined in \autoref{sec:sigma8}. Years of particular interest for cosmological analyses are indicated with vertical lines, where the uniform rolling strategies (\vfpz\ and \vtpfur) reach the lowest value of $\sim$0.5, which corresponds to fluctuations of depth in strategies without rolling features.}
    \label{fig:sig8metric}
\end{figure}

The results for the fiducial strategies analyzed in this work are shown in \autoref{fig:sig8metric}.
Vertical years indicate the key years at which static cosmological analyses would be carried out, when uniformity is critical. 
We now discuss the three main features seen in this figure: the large bias for Year 1; the stabilization of the bias (for all strategies) at $0.5-1\sigma$ for subsequent years; and the ability of the uniform rolling strategy to recover near-uniformity at key years. 

The large bias for Year 1 is the result of the highest redshift bin being the most affected by depth fluctuations and also carrying the most weight due to the higher number of modes in the angular power spectra, see \autoref{fig:sig8_nz}.
This effect decreases in subsequent years as the depth increases and the highest redshift bin becomes less sensitive to non-uniformity.
Importantly, these results assume a fixed extragalactic analysis mask only including a cut based on Galactic dust extinction, which facilitates the interpretation of yearly changes.
However, in real analyses one would adjust the mask in order to cut areas not meeting a sufficient depth in all bands (which would mitigate the effect above). We have explicitly confirmed that applying such a cut significantly improves the result in \irtwo{Y}ear 1 while having little effect in later years.  Less trivially, the redshift bins and angular scales considered in the analysis could also be adapted based on data quality verification or simulations. 
We defer this to future work.

However, we note that adjustments such as sky or scale cuts cannot fully mitigate the effect of non-uniformity.
This is in part demonstrated by the fact that years after Year 1 exhibit depth variations of \cwr{$\lesssim 1$} magnitude, which would not warrant significant sky cuts, but the combined effect results in a $\sim 0.5\sigma$ bias on $\sigma_8$.
This level corresponds to the typical bias resulting from the current levels of near-uniformity achieved in these simulations due to weather conditions.
It is the value that strategies with no rolling achieve; 
however, these strategies are no longer realistically considered, as rolling will be implemented in the LSST. 
We see that the \vtpfur\ and the \vfpz\ (which both have uniform rolling) are consistent with no rolling at the key years of interest.  They show a $20-40$\% improvement relative to \vtpf\  and \vfpzfc\, while matching them in other years. 

We now discuss the assumptions of this metric in more detail, and how they may affect the results.
Its main parameters (other than the derivatives of the linear contamination), are the galaxy bias and the fraction of uncorrected power, which are fixed to unity and 10\% \ir{as motivated above}, respectively.
They affect the absolute values of the metric, making them non-trivial to interpret.  
However, given that both factors introduce  multiplicative factors on the power spectra, the {\em relative} $\sigma_8$ bias between strategies and years is not very sensitive to these numbers, and our conclusions are robust. 
Other approximations made in this metric include the Gaussian covariance matrix and the scale cuts $\ell_\mathrm{max}=k_\mathrm{max}\chi(z)$, taken at the mean redshift of each tomographic bin.
These issues can be ignored for our analysis since they only complicate the modeling and will have a small effect on the inferred bias on $\sigma_8$, leaving the relative differences between strategies unchanged.

The main limitation of this metric is the realism in the modeling of the spurious fluctuations, for two reasons.
First, linear contamination is correctable with existing methods \cite[e.g.,][]{2021MNRAS.503.5061W,2024arXiv240112293B,2020JCAP...03..044N}. 
Second, real contamination is non-linear \citep[see e.g.,][]{Elvin_Poole_2018, Rodr_guez_Monroy_2022, Nicola_2020}.
Some of the density fluctuations, if they are due to depth but not to other sources of systematics (such as stellar contamination and blending, which are non-linearly coupled to depth), may be correctable with a framework such as that of \cite{baleato/etal:2023}.
However this requires accurate modeling or calibration of the spatially-varying redshift distributions, which has not been demonstrated in practice.
Nevertheless, we believe that 10\% of our linear model should give an indication of the uncorrected spurious power for a \ir{magnitude-limited} galaxy sample. 
Moreover, conclusions based on the relative differences between strategies should be even less sensitive to the modeling assumptions. 

\subsection{Metric: mean $z$ (\texttt{MultibandMeanzBiasMetric})} 
\label{sec:meanz}

Using the derivative in the mean redshift of the galaxy sample and the dependence of the fluctuations in the galaxy distribution as a function of depth, we can convert fluctuations in depth across the survey (captured through the RMS scatter in the depth for each band) into changes in the mean redshift of the sample, as described in \autoref{sec:simulation}.

These fluctuations map directly onto a bias in the recovered cosmic shear signal \citep{baleato/etal:2023}, through the coupling of the covariance of the perturbations across a comoving distance slice and the lensing shear power spectrum. Here the convergence field can be constructed from the overdensity field $\delta$ by projecting along the line of sight:
\begin{equation}
\kappa_{\ell m} = \int \mathrm{d}\chi \left[\bar{g}(\chi)\delta_{\ell m}(\chi) + \{\Delta g \, \delta\}_{\ell m}(\chi) \right],
\end{equation}
where \begin{equation}
   \bar{g}(\chi) \equiv \frac{3}{2}\Omega_m\frac{H^2_0}{c}\frac{\chi}{a(\chi)}\int_0^{\infty} \mathrm{d}\chi_s\frac{H(\chi_s)}{c}\frac{(\chi_s-\chi)}{\chi_s}\frac{\mathrm{d}\bar{n}_g}{\mathrm{d}z}(\chi_s)
\end{equation}
is the lensing kernel given a distribution of sources $\mathrm{d}n_g/\mathrm{d}z$.  Quantities with a bar on top are spatial averages, while $\{\Delta g\,\delta\}_{\ell m}$ encodes the spatial variations in the galaxy number density of source galaxies, which affects the number of measurements available to extract the shear signal, imprinting
inhomogeneity in the shape noise across the sky. $\chi$ is the comoving distance.  Following \citealt{baleato/etal:2023}, we assume a flat universe and work in the Born approximation.

With simplifying assumptions (e.g., the Limber approximation), the fluctuations in the angular cross spectrum between two tomographic redshift bins $i$ and $j$ can be written as:
\begin{equation}
T_\ell^{ij} \simeq \int \frac{\mathrm{d}\chi}{\chi^2}\left[\left\{\bar{g}^i(\chi)\bar{g}^j(\chi)+\frac{\mathrm{Cov}[g^i,g^j](\chi)}{\chi^2}\right\}P_{ij}\left(\frac{\ell+1/2}{\chi};z\right)\right], \label{eq:clbias}
\end{equation}
where $P_{ij}$ is the 3D galaxy power spectrum and the covariance $\mathrm{Cov}[g^i,g^j](\chi)$ is related to the angular power spectrum
\begin{equation}
    \mathrm{Cov}[g^i,g^j](\chi) = \mathrm{Cov}[\Delta g^i,\Delta g^j](\chi) = \sum_\ell \frac{(2\ell +1)}{4\pi}C^{\Delta g^i\Delta g^j}(\chi).
\end{equation}
Given the simulated fluctuations in the mean redshift and the densities described in \autoref{sec:simulation}, we calculate the bias in the shear power spectrum, interpolating over the bias values shown in Figure 9 in \cite{baleato/etal:2023}, which is related to \autoref{eq:clbias}. 
We do so by taking the mean redshift of each redshift bin and taking the mean of the biases over the 5 redshift bins \ir{to obtain the \texttt{MultibandMeanzBiasMetric}. This metric can be seen as a measure of bias in the estimated weak lensing signal from changes to the redshift distribution of the lensing sample, analogous to the \texttt{TomographicClusteringSigma8biasMetric} bias in the clustering signal described in \autoref{sec:sigma8}.} 

\begin{figure}[htbp!]
\centering
\begin{tabular}{cc}
\includegraphics[width=0.9\textwidth,trim={0.1cm 0.1cm 0.1cm 0.1cm},clip]{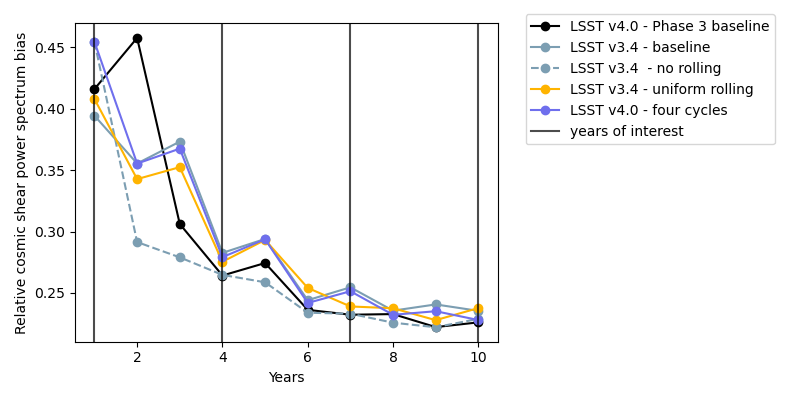}
    \end{tabular}
    \caption{This shows the \texttt{MultibandMeanzBiasMetric} defined in \autoref{sec:meanz}, corresponding to the bias in cosmic shear angular power spectra (\cwr{a}veraged over the \cwr{5} redshift bins) arising from the combined depth fluctuations in the $riz$ bands. 
    \label{fig:meanz}}
\end{figure}

In the DESC Science Requirements Document \citep{DESCSRD}, the \cwr{requirement} on the systematic uncertainty in the redshift-dependent shear calibration is 0.013 in the Y1 analysis and 0.003  in the Y10 analysis. In \autoref{fig:meanz} we show the ratio of the bias in the shear relative to the Y10 requirement. Functionally, this calculation makes the assumption that we are not modeling the impact of spatial variation in the mean sample redshift in the weak lensing shear data vectors, so their contribution is effectively a bias in the signals (with a different amplitude at different redshifts).  \cwr{As shown, the uniform rolling approach comes close to the case without rolling at the years of interest, and even without rolling the effect may be large enough to require modeling.}

\section{
Conclusions}\label{sec:disc/conc} 

\cwr{The key problem motivating this work is the spatially-varying survey depth of LSST (or `non-uniformity') at intermediate survey years imprinting large-scale patterns in the data that affect static cosmological probes such as weak lensing and galaxy clustering. Survey non-uniformity has been \ir{demonstrated} to be a limiting factor in current (Stage III) surveys \citep{Heymans_2021, yan2024kidslegacyangulargalaxyclustering, Elvin_Poole_2018, Andrade_Oliveira_2021, Rodr_guez_Monroy_2022, Nicola_2020, miyatake2023hypersuprimecamyear3, Sugiyama_2023} and promises to be a challenge for the \ir{full utilization} of the LSST, in particular because of the potentially large survey non-uniformity arising due to the rolling phases included in the observing strategy to enhance transient science \citepalias{scoc_phase2}, along with the large dataset size and correspondingly small statistical errors and systematic tolerances.} 

\cwr{
In this work, we calculated the effects of imaging depth upon galaxy number density and mean redshift.  We then developed and 
explored new metrics to quantify the resulting effect of non-uniformity of the LSST imaging depth on static cosmological analyses.
These metrics included the stripiness of combined $r+i+z$ exposure time, the area placed at risk for cosmological analysis by non-uniform imaging depth, the bias in measuring the cosmological parameter $\sigma_8$ produced by non-uniform depth, and the bias in shear measurements caused by changes in the source redshift distribution.  For the previous implementation of rolling cadence, we found that nearly half of the survey could be lost for static science at intermediate data releases, decreasing the Dark Energy figure of merit by approximately 40\%.  
}

\cwr{\ir{As further motivation to update the observing strategy, we found that at years between 1 and 10 in the \vtpf\ simulation, the rolling stripe features were so prominent that the survey would have to effectively be divided into two different (yet interleaved) surveys that could not be analyzed jointly with current methodology or straightforward variants thereof.} The metrics developed in this paper informed the development of a new observing strategy (coined `uniform rolling') that not only includes phases of rolling but also re-establishes near-uniformity at specific times of interest (years 4 and 7) to facilitate static cosmological analyses \ir{between year 1 and year 10}.
This was originally demonstrated on the v3.4 simulated strategies and presented to the Rubin SCOC in June 2024, which motivated the SCOC to adopt the uniform rolling strategy as the baseline for \ir{the v4 survey strategy} \citepalias{scoc_phase3}.}

This paper verified on the latest set of simulated strategies (v4.0), which now have uniform rolling \ir{included in the baseline strategy (\vfpz), that the rolling stripe features are not significant at years 4 and 7, facilitating a more straightforward static science analysis at those intermediate years.  We}  also demonstrated that \ir{strategies with} four rolling cycles \ir{as currently implemented} cannot return to uniformity at years 4 and 7, as shown by considering \vfpzfc. \ir{In all cases, at least one year of uniform observations was included at the start of the survey \cwr{since that is the default for all strategies under consideration at this time.}}
Future work will include the development of new metrics, which could prove important for tracking uniformity during the LSST and informing adjustments to the observing strategy if necessary.

\clearpage
\section*{Author contribution statements}
Author contributions are listed below.\\
B. Leistedt: Analysis, metric development, participation in project discussions. Writing - original draft, writing - review and editing. 
M.~R.~Becker: Led the development and technical implementation of the uniform rolling survey strategy.
\cwr{H. Awan: Participation in project discussions, internal review.}
E.~Gawiser:  Contributed ideas to development of uniform rolling strategy, writing and editing.  
Q.~Hang: Making the mock galaxy catalogue for some metrics, writing and editing.
R. Hlo\v{z}ek: Analysis, metric development, participation in project discussions. Writing - original draft, writing - review \& editing.
\cwr{S. Jha: Internal reviewer, builder (contributions to survey strategy, rolling cadence, etc.).}
\cwr{L. Jones: Supported development of uniform rolling strategy.}
\cwr{A. Kannawadi: Contributed to early analyses identifying the impact of limiting magnitudes on mean photometric redshifts.}
\cwr{M. Lochner: Builder (significant contributions to observing strategy optimization over several years), reviewed paper.}
R.~Mandelbaum: Analysis, metric development, participation in project discussions, writing and editing. 
J. Newman: Analysis, metric development, writing.
\cwr{I. Sevilla-Noarbe: Internal review of paper.}
\cwr{H. Peiris: Builder (significant contributions to observing strategy optimization over several years), DESC liaison in SCOC (direct contribution to rolling cadence transient metrics and facilitating homogeneity discussions).}
\cwr{E. Rykoff: Contributed to discussion of metrics and impact of uniformity on photometric calibration.}
\cwr{M. Troxel: Contributed to discussion and formulation of metrics.}
P.~Yoachim: Code implementation of uniform rolling and simulations of observing strategy.

\begin{acknowledgments}
\section*{Acknowledgements}
BL is supported by the Royal Society through a University Research Fellowship.  RM is supported by the Department of Energy Cosmic Frontier program, grant DE-SC0010118. RH is supported by the Canadian National Science and Engineering Research Council Discovery Grant RGPIN-2025-06483. ML acknowledges support from the South African Radio Astronomy Observatory and the National Research Foundation (NRF) towards this research. Opinions expressed and conclusions arrived at, are those of the authors and are not necessarily to be attributed to the NRF. HVP was supported by funding from the European Research Council (ERC) under the European Union’s Horizon 2020 research and innovation programmes (grant agreement no.~101018897 CosmicExplorer).

This paper has undergone internal review by
the LSST Dark Energy Science Collaboration. The internal reviewers were Humna Awan, Saurabh Jha and Nacho Sevilla.
The authors thank the internal reviewers for their valuable comments\cwr{, along with feedback received during collaboration-wide review}.
The DESC acknowledges ongoing support from the Institut National de 
Physique Nucl\'eaire et de Physique des Particules in France; the 
Science \& Technology Facilities Council in the United Kingdom; and the
Department of Energy and the LSST Discovery Alliance
in the United States.  DESC uses resources of the IN2P3 
Computing Center (CC-IN2P3--Lyon/Villeurbanne - France) funded by the 
Centre National de la Recherche Scientifique; the National Energy 
Research Scientific Computing Center, a DOE Office of Science User 
Facility supported by the Office of Science of the U.S.\ Department of
Energy under Contract No.\ DE-AC02-05CH11231; STFC DiRAC HPC Facilities, 
funded by UK BEIS National E-infrastructure capital grants; and the UK 
particle physics grid, supported by the GridPP Collaboration.  This 
work was performed in part under DOE Contract DE-AC02-76SF00515.

Part of this work was completed on Hyak, UW’s high performance computing cluster. This resource was funded by the UW student technology fee.
\end{acknowledgments}

%

\vspace{5mm}


\software{Astropy \citep{astropy2013, astropy2018}, Healpy and HEALPix\footnote{\url{http://healpix.sourceforge.net}} \citep{healpix2005, Zonca2019}, Numpy \citep{harris2020array}, Matplotlib \citep{Hunter:2007}, Scipy \citep{2020SciPy-NMeth}, Pandas \citep{reback2022}, rubin\_scheduler \citep{rubin_scheduler2024}, 
PhotErr \citep{2024AJ....168...80C},
rubin\_sim \citep{rubin_sim2024}, conda-forge \citep{conda_forge}}



\appendix

\section{Uniform Rolling}
\label{app:uroll}

\begin{figure*}
    \centering
\includegraphics[width=\textwidth]{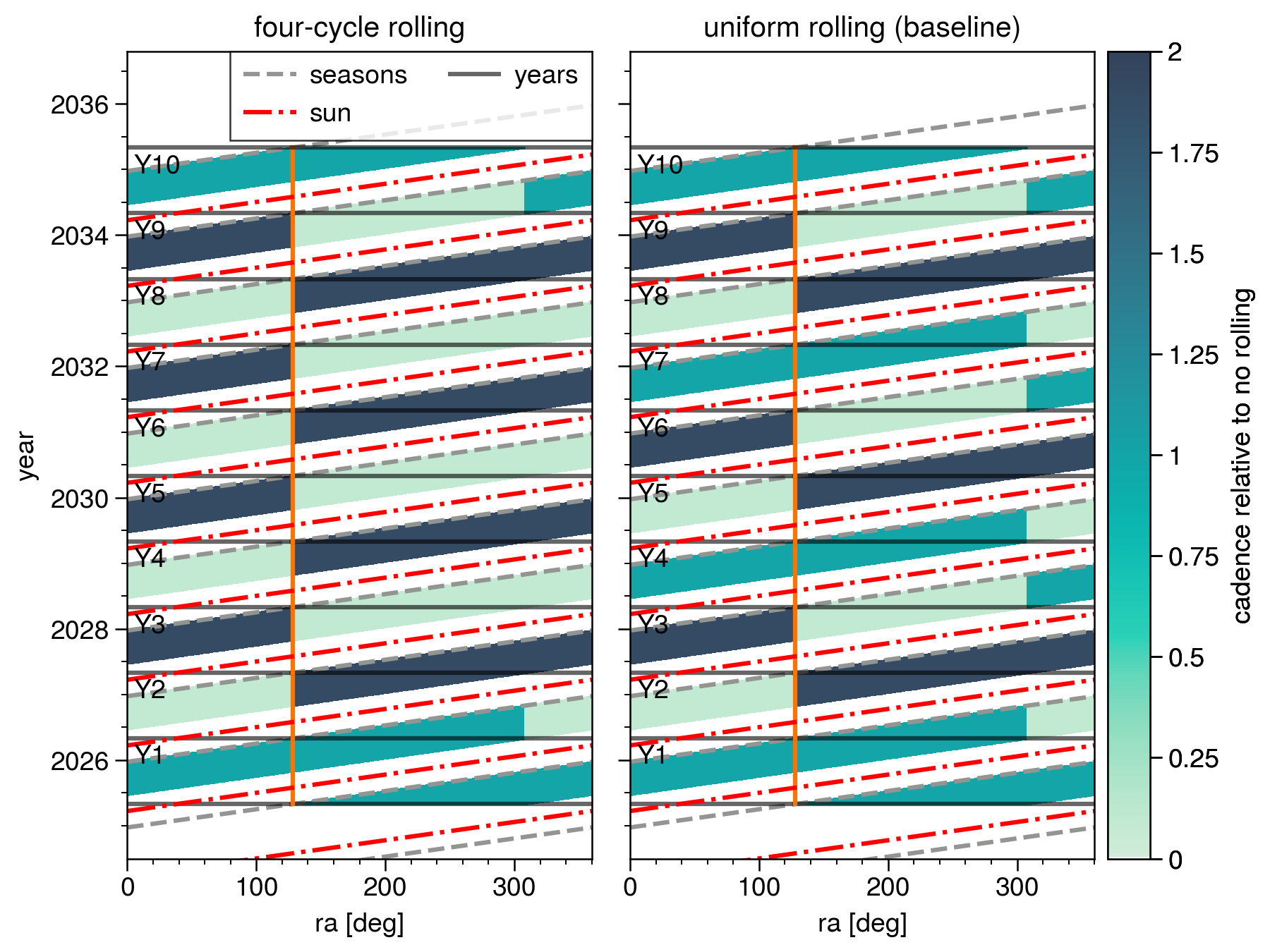}
    \caption{
      Rolling strength (or overall cadence with one being the nominal \noroll\ cadence) in seasons as a function of time and right ascension for half of the sky in a two-stripe rolling strategy. The other half of the sky has the seasons of low- and high-cadence swapped.   \ir{See \autoref{app:uroll} for a full explanation of this figure\cwr{, including descriptions of the position of the sun and the season starts in red and gray respectively as shown in the legend.}}  
      \ir{We note that in the \cwr{baseline  rolling strategy (\vtpf)}  a specific start time of the rolling was assumed, however future strategies may vary this (for example they may start rolling at the same time as \cwr{\vtpfur)}. The time at which rolling begins may also alter survey uniformity.}}
    
    \label{fig:rolling}
\end{figure*}

In this appendix, we provide more details on the implementation of ``uniform'' rolling. \cwr{In \autoref{fig:rolling} we illustrate} these details
and forms the basis of the discussion here. This figure shows as a function of right ascension (horizontal axis) and time (vertical axis), the rolling
strength for one of the two sets of stripes in declination in the sky. \cwr{Recall that rolling works by alternating low- vs high-cadence seasons and that these have to be balanced in declination in order to maintain the overall cadence of the survey. At a given time and location in right ascension, we can only show one of two regions in our two-dimensional plot. The other region has the cadences reversed in order to balance the survey.} The left plot in this figure shows the \cwr{\vtpf. The right plot shows the \vtpfur\ survey.}

We have marked several other important features \cwr{i}n this plot. First, the season starting times for each location in right ascension are shown as the gray dashed lines. This line is sloped since not every part of the sky can be in season at once. If \cwr{one picks} a location in right ascension and follow it vertically, \cwr{one can} read off the patterns of low-, high-, and average cadence. Second, when a location is ``in season'' (i.e., it is in the last six months of its season), we show the cadence using the color bar varying from the lowest cadence, 0.1, to the highest cadence, 1.9. These numbers mean a location gets only 10\% of the observations it would get in a ``noroll'' survey (i.e., 1.0) or 90\% more observations (i.e., 1.9). Each calendar year of the survey is shown as the horizontal \irtwo{solid} lines. The red dot-dashed lines show the position of the sun in right ascension and time. As is readily apparent, the scheduler does not point the telescope at the sun. Finally, \cwr{the vertical orange} line shows 90 degrees plus the right ascension of the position of the sun at the nominal survey start date. This value of right ascension plays a special role in uniform rolling as it is used to break up the sky in right ascension in order to interleave the survey's seasons.

With these basic descriptions, we can examine a uniform rolling cycle by looking at the right panel of \autoref{fig:rolling} and focusing on the time period from Year 1 to Year 4. During this period the survey executes a series of cadences as a function of right ascension. It is most easy to follow the survey as a function of season (i.e., move diagonally in the figure), where it uses a pattern of average cadence (1.0 for the partial seasons), low cadence (0.1 for half of the right ascension range), high cadence (1.9 for the full right ascension range), low cadence (0.1 for half of the right ascension range), average cadence (1.0 for half of the right ascension range), and finally average cadence for the other half of the partial season at the start. \cwr{By partial season, we mean that a season of high cadence crosses a horizontal year line on the plot.}  This pattern results in every location in right ascension getting three full seasons, one each of average cadence, low cadence, and high cadence, over the full three-year time period with no partial seasons of any low or high cadence rate crossing the year four time cutoff. This feature is the key to uniform rolling. One needs the additional average cadence season wait for other parts of the sky to finish their rolling cycle and to account for the partial season encounter\cwr{ed} in the first six months after year one. If instead, one started a new rolling cycle again, it would not be possible to finish it, leaving the survey with large depth variations. To build up the full survey strategy, we stack three of these three-year cycles on top of an initial full season of average cadence for the whole sky to form the full ten-year survey.

It is useful to compare to the \vtpf\ survey \cwr{i}n the left panel of this figure. Here, following right ascension vertically from years one to four, it is evident that not all parts of the survey execute three full seasons with one each of average cadence, low cadence and high cadence. Instead, some parts execute a sequence that is high cadence, low cadence, and then high cadence again in a partial season between years one and four. 
The \vtpf\  survey starts rolling right away after each cycle is done and so leaves incomplete rolling cycles at any given time (horizontal lines).

\begin{figure*}
    \centering
    \includegraphics[width=\textwidth]{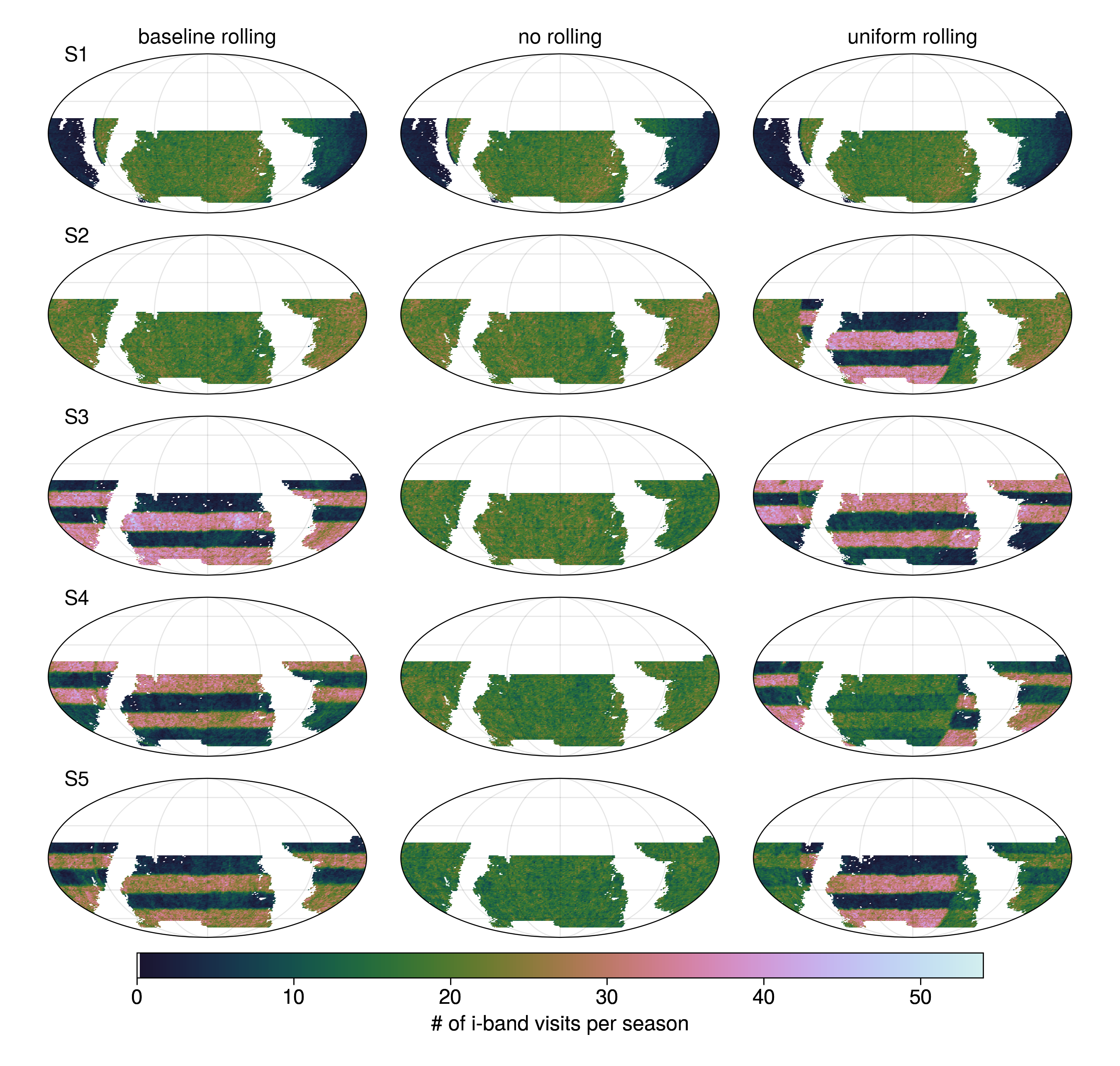}
    \caption{The number of visits per season in the $i$ band for first five seasons in the \vtpf, \noroll, and \vtpfur\ cases. Each season is plotted on each row and the various strategies are shown in the columns. The \vtpf\ strategy executes 1.5 rolling cycles, with a complete cycle executed in seasons 3 through 4. The \vtpfur\ strategy takes an additional season to execute a full rolling cycle, using seasons 2 through 4. Season 5 is the start of the next rolling cycle for both strategies. The ``ghosted'' stripes in the uniform rolling seasons 4 and 5 maps (e.g., at the center of season 4) are due to the scheduler working to reach a total cumulative number of visits and thus compensating for depth variations imparted in previous seasons. 
    }
    \label{fig:rollingvisits}
\end{figure*}

Another way to examine these differences is to directly look at the number of visits per season for each location on the sky. This information is shown in \autoref{fig:rollingvisits}. Each row shows a season (e.g., the first row shows season one denoted as \verb|S1|) and each column shows one of the three survey strategies (i.e., \vtpf, \noroll\ and \vtpfur). For the \vtpf\ rolling cadence, 1.5 cycles of rolling cover seasons 3-5 with one cycle being seasons 3 and 4. For the \vtpfur\  cadence, a single rolling cycle is executed from seasons 2 through 4. Season 5 is the start of the next uniform rolling cycle. Further, for the \vtpfur\ case, the cadence of alternating rolling stripes is broken up both by season and by right ascension. As described above, this ensures uniform coverage at specific calendar times. Finally, seasons 4 and 5 for the \vtpfur\ cadence show an interesting feature of this strategy. Even in regions of right ascension where the scheduler is told to execute a uniform survey (e.g., the center portion of season 4), the rolling stripes are evident. These ``ghosted'' stripes occur because the scheduler works towards a total cumulative number of visits. If the scheduler sees that a given area is too far ahead or behind the total number of expected visits after the previous rolling season, it uses the uniform season in that part of the sky to even out the survey to reach a uniform cumulative total number of visits.

\section{Derivation of the tomographic $\sigma_8$ metric}
\label{appsigma8}

To derive the \texttt{TomographicClusteringSigma8biasMetric}, we first consider the linear matter power spectrum $P(k, z)$ and recall that it is exactly proportional to $\sigma_8^2$, by definition.
The dependency on other cosmological parameters is neglected, i.e., they are assumed to be fixed to fiducial values.
This allows us to derive an analytic expression for the bias in $\sigma_8^2$ when depth fluctuations give rise to density fluctuations which are uncorrected for.

Under the Limber approximation, the angular power spectrum of galaxy overdensities in a redshift bin indexed by $i$ reads 
$$
C_\ell^{g_i g_i} = \int  \frac{\mathrm{d}\chi}{\chi^2} W_i^2(\chi) P\left(k=\frac{\ell+1/2}{\chi}, z\right)
$$
with $W_i(\chi)$ a kernel containing the redshift distribution and galaxy bias.
The total variance of the density fluctuations is
$$
\sum_{\ell = \ell_\mathrm{min}}^{\ell_\mathrm{max}(i)} (2\ell+1) C_\ell^{g_i g_i}  \propto \sigma_8^{2}
$$
We use a fiducial cosmology $(\Omega_c=0.25, \Omega_b=0.05, h=0.7, n_s=0.95, \sigma_8=0.8)$
and compute the angular power spectra with CCL \citep{2019ApJS..242....2C} for 5 tomographic redshift bins following the redshift distributions from the simulations presented in \autoref{sec:derivatives}.
The input redshift distributions and the calculated angular power spectra are shown in \autoref{fig:sig8_nz}.
We set the galaxy bias to unity.
Furthermore, we consider scales large enough so that the Poisson noise can be neglected.

The next step in this derivation is to leverage the global proportionality to $\sigma_8^2$ setup in tomographic redshift bins by defining some effective transfer functions 
$$
T_i \equiv \sigma_8^{-2} \sum_{\ell = \ell_\mathrm{min}}^{\ell_\mathrm{max}(i)} (2\ell+1) C_\ell^{g_i g_i}
$$
where the sum over $\ell$ needs to cover the same (linear) scales between redshift bins. From the limber approximation we can set $\ell_\mathrm{max}(i) = k_\mathrm{max} \chi(z_i)$  with $\chi(z_i)$ the comoving radial distance at the redshift $z_i$, which would be the peak of the redshift distribution of that bin. This neglects interlopers, but this assumptions simplifies greatly this analysis and allows us to make clear scale cuts and to conserve the linearity in $\sigma_8^{2}$.

The simplest approach on the observations side is to assume that we measure $\hat{C}_\ell^{g_i g_i}$ as a Gaussian draw with mean $C_\ell^{g_i g_i}$ and (cosmic) variance $V_\ell^{g_i} = 2 C_\ell^{g_i g_i} / f_\mathrm{sky} / (2\ell+1)$ with $f_\mathrm{sky}$ the sky coverage. Furthermore, we can define the observed integrated angular power spectra as
$$
\hat{y}_i = \sum_{\ell = \ell_\mathrm{min}}^{\ell_\mathrm{max}(i)} (2\ell+1) \hat{C}_\ell^{g_i g_i}
$$
which correspond to the total variance of the field measured in each redshift bin (integrated over linear scales).
$\hat{y}_i$ itself has a variance $V_i = \sum_{\ell = \ell_\mathrm{min}}^{\ell_\mathrm{max}(i)} (2\ell+1)^2 V_\ell^{g_i}$.

We now derive an estimator for $\sigma_8$, leveraging the fact that $\sigma_8^2$ is a common multiplicative term between all values $\hat{y}_i/T_i$, and that each $\hat{y}_i$ is Gaussian distributed with a variance $V_i$.
Thanks to the properties of Gaussian distributions \citep{Petersen2008}, we can derive a posterior distribution on $\sigma_8^2$ which itself is Gaussian \citep{hogg2020dataanalysisrecipesproducts}. It has a mean of $\hat{\sigma}_8^2 = F_\mathrm{OT}/F_\mathrm{TT}$ and a variance $1/F_\mathrm{TT}$, with
$F_\mathrm{OT} = \sum_i \hat{y}_i T_i / V_i $ and
$F_\mathrm{TT} = \sum_i T_i T_i / V_i $.
This assumes a uniform prior on $\sigma_8^2$ (although a Gaussian prior could be included).

We now apply this idea to density fields which include some level of contamination due to spatially varying depth.
To do so, we first recall that if a density field (in the $i$th redshift bin) is additively contaminated by a field $c_i$, i.e., if $\delta^\mathrm{obs}_{i}(\vec{n}) = \delta^\mathrm{true}_{i}(\vec{n}) + c_i(\vec{n})$, then the result on the angular power spectra is also an additive contamination (by the angular power spectra of the contamination fields $c_i$).
This is also a good approximation for multiplicative contamination on large scales \citep{Huterer2013_lsssystematics, Shafer2015_lsssystematics}.
If we have a model for $c_i$ (e.g., from MAF maps) for each tomographic bin, then we can measure the angular power spectra $\hat{C}_\ell^{c_i c_i}$, and define the observed total variances as
$$
\hat{y}_i = \sigma_8^2 T_i + f  \sum_{\ell = 1}^{\ell_\mathrm{max}(i)} (2\ell+1) \hat{C}_\ell^{c_i c_i}
$$
where $f \in [0, 1]$ controls the fraction of the power which is uncorrected.
We introduce this parameter because it is likely that some of the spurious fluctuations can be corrected by mitigation techniques.
For example, linear mitigation techniques based on contamination templates can exactly cancel out the spurious contamination which is linear in the templates \citep[e.g.,][]{2017MNRAS.465.1847E, 2021MNRAS.503.5061W}.
This is the case of our model.
We assume $f=0.1$, which \ir{is a reasonable heuristic representing the typical efficacy of clustering systematics mitigation as discussed in \autoref{sec:sigma8}.}
This simple definition allows us to avoid simulating the underlying cosmological fields and to measure the level of contamination directly\ir{, which would require introduction of other assumptions about the types and levels of imaging systematics in Rubin data}.
As a result, our derived posterior distribution will have a biased $\hat{\sigma}_8^2$ which we can compute with the equations above. Thus, $(\hat{\sigma}_8^2 -
\sigma_8^2)\sqrt{F_\mathrm{TT}}$ corresponds to this bias in units of the standard deviation. With no contamination, we would recover $\hat{\sigma}_8^2=\sigma_8^2$.
The bias and variance in $\sigma_8^2$ can be converted into corresponding values for $\sigma_8$ using simple propagation of uncertainties.

          
\bibliography{sample631}{}
\bibliographystyle{aasjournal}



\end{document}

%% file: simulation.tex
We construct mock LSST photometry catalogues using the Roman-Rubin simulation \cwr{(\citealt{2025arXiv250105632O}, specifically \texttt{roman\_rubin\_2023\_v1.1.3\_elais}\footnote{\url{https://roman.ipac.caltech.edu/page/nasa-openuniverse-images-2024}})}. 
This simulation is based on the `Outer Rim' N-body cosmological simulation \citep{2019ApJS..245...16H}.
Galaxies are simulated by first building a parametric model that links galaxy star formation history with physical parameters in halo mass assembly, using Diffstar \citep{2023MNRAS.518..562A}, and then assigning its SED as a function of its star formation history, also taking into account other properties such as its metallicity and dust, using Differentiable Stellar Population Synthesis \citep[DSPS; ][]{2023MNRAS.521.1741H}. With this separate modeling for different galaxy components, the Roman-Rubin simulation is able to produce realistic fluxes and colours in the six LSST bands, enabling a more realistic investigation of depth variation on photo-$z$.\footnote{One should note that the current version of the Roman-Rubin simulation has an issue at $z>1.5$. There is an obvious bimodal distribution in the $g-r$ versus redshift plane, which is not observed in the real galaxy data. This could be due to the fact that these high redshift observations are not as constraining when fitted with the SPS models. This feature could affect the performance of the photo-$z$ estimator. For example, a machine-learning algorithm may be able to achieve better results than expected based on the object's $g-r$ colours. Therefore, one should be careful in interpreting the results from high-redshift samples.}

The truth sample\ir{, meant to represent a potentially observable galaxy sample without any photometric noise or other observational effects,} is constructed by randomly selecting $10^6$ objects from the Roman-Rubin simulation, complete to $i<26.5$. 
In addition to the ${ugrizy}$ magnitudes and true redshifts, we also compute the galaxy semi major- and minor axes $a,b$, in order to compute the extended magnitude errors for the catalogue (elaborated below). Given the bulge size $s_b$, disk sizes $s_d$, bulge-to-total ratio $f_b$, and ellipticity $e$ of the galaxy, the semi major- and minor axes are: $a=s/\sqrt{q}$ and $b= s\sqrt{q}$, where $s$ \cwr{is} the weighted size of the galaxy, $s= s_bf_b + s_d(1-f_b)$, and $q$ is the ratio between the major and minor axis, related to ellipticity via $q=(1-e)/(1+e)$. 

\cwr{We randomly assign these million galaxies to positions in the Rubin WFD footprint and degrade them accordingly.} Specifically, we degrade the truth sample \ir{to include the potential effects of noise in the photometric observations} by first computing the expected photometric error for each galaxy for each band using the LSST error model \citep{2019ApJ...873..111I}. 
{The noise-to-signal ratio of the flux consists of two components:}
\begin{equation}
    {\sigma}^2 = {\sigma}_{\rm sys}^2+ {\sigma}_{\rm rand}^2,
    \label{eq: mag err 1}
\end{equation}
where $\sigma_{\rm sys}$ is the irreducible error of the system in AB magnitudes, set to 0.005 as in \cite{2019ApJ...873..111I}, and $\sigma_{\rm rand}$ is the random error. {The photometric error is given by $\sigma_m = 2.5\log_{10}(1+\sigma)$. In the high signal-to-noise case, $\sigma_m\approx\sigma$ \citep{2019ApJ...873..111I}. }
For point-like objects, $\sigma_{\rm rand, pt}$ is given by
\begin{equation}
    {\sigma}_{\rm rand, pt}^2=(0.04-\gamma)x+\gamma x^2, 
    \label{eq: mag err 2}
\end{equation}
where $\gamma$ is a band-dependent parameter as in \cite{2019ApJ...873..111I}, and $\log_{10} x \equiv 0.4 \left(m-m_5\right)$. Here, $m$ is the magnitude of the object in a specific band, and $m_5$ is the point-like $5\sigma$ depth in that band. For extended objects, the random error receives an extra factor, following \cite{2020A&A...642A.200V,2019A&A...625A...2K}:
\begin{equation}
    {\sigma}_{\rm rand} = {\sigma}_{\rm rand, pt} \sqrt{{A_{\rm ap}}/{A_{\rm psf}}},
    \label{eq: extended err}
\end{equation} 
where $A_{\rm psf}$ is the PSF size given by
\begin{equation}
    A_{\rm psf}=\pi \sigma_{\rm psf}^2, \quad \sigma_{\rm psf}=\theta_{\rm FWHM}X^{0.6}/2.355. 
\end{equation}
Here, $\theta_{\rm FWHM}$ is the full-width half maximum seeing in arcsec, $X$ is the airmass, and $X^{0.6}$ accounts for the increase in seeing with airmass (but not the increase in PSF size due to differential chromatic refraction). 
\begin{equation}
    A_{\rm ap}=\pi a_{\rm ap}b_{\rm ap},\quad
    a_{\rm ap}  = \sqrt{\sigma_{\rm psf}^2+(2.5a)^2}, \quad
    b_{\rm ap}  = \sqrt{\sigma_{\rm psf}^2+(2.5b)^2},
\end{equation}
where $a, b$ are the galaxy semi-major and minor axes in arcsec. 

The \ir{magnitudes from the truth sample are} then degraded in the following way. In each band, the magnitude is first converted into flux via $f=10^{-m/2.5}$. Then, a flux error $\Delta f$ is drawn from a normal distribution with width $\sigma$: $\Delta f\sim \mathcal{N}(0,{\sigma})$. Finally, the degraded magnitude is obtained by converting the degraded flux back to magnitudes: $m_{\rm deg}=-2.5\log_{10}(f+\Delta f)$. Again, in the high signal-to-noise limit where $\Delta f \ll f$, $\Delta f\approx \Delta m$, and one recovers the expression in \cite{2019ApJ...873..111I}.
Negative fluxes are set as `non-detection' in that band.

\begin{table}
\centering
 \caption{Parameters used in \texttt{photerr} in order to compute the magnitude error. $\gamma$ is a band-dependent parameter  \citep{2019ApJ...873..111I}, $\theta_{\rm FWHM}$ is the full-width half maximum seeing in arcsec, $X$ is the airmass, and $m_5^{\rm Y10}$ is the $5\sigma$ coadd depth for LSST 10-year observation.}
 \label{tab: lsst error model params}
 \begin{tabular}{lcccc}
  \hline
  & $\gamma$ & $\theta_{\rm FWHM}$ & $X$ & $m_5^{{\rm Y10}}$\\
  \hline
  $u$ & 0.038 & 1.22 & 1.15 & 25.6\\
  $g$ & 0.039 & 1.09 & 1.15 & 26.9\\
  $r$ & 0.039 & 1.02 & 1.14 & 26.9\\
  $i$ & 0.039 & 0.99 & 1.15 & 26.4\\
  $z$ & 0.039 & 1.01 & 1.16 & 25.6\\
  $y$ & 0.039 & 0.99 & 1.16 & 24.8\\
  \hline
 \end{tabular}
\end{table}
The above procedures are carried out using the python package \texttt{photerr}\footnote{\url{https://github.com/jfcrenshaw/photerr}} v1.1.1 \citep{2024AJ....168...80C}. The LSST error model parameters used are listed in Table~\ref{tab: lsst error model params}. 
We interpolate the `nominal' $5\sigma$ depth for each year of observation via $m_5^{\rm nom}(N_{\rm yr})=m_5^{\rm Y10}+2.5\log_{10}(\sqrt{N_{\rm yr}/10})$, where $m_5^{\rm Y10}$ is the nominal depth for 10-year observation, and is listed in the last column in Table~\ref{tab: lsst error model params} for each band.
To systematically study the shifts in number densities and redshifts in the resulting galaxy samples, we construct a degraded sample, for each year, by perturbing around the nominal depth one band at a time.
The perturbed depth has ten bins in $[m_5^{\rm nom}-0.5, m_5^{\rm nom}+0.5]$ with a bin size of 0.1 mag. Depth of other bands are fixed to the nominal value. 
We also include a case where the depth for all six bands are varied at the same time. \ir{Finally, for each sample, we apply a limiting magnitude cut of $i<i_{\rm lim}^{\rm Y10}=25.3$ \citep{DESCSRD}, and we adjust it for previous years by $i_{\rm lim}(N_{\rm yr})=i_{\rm lim}^{\rm Y10}+2.5\log_{10}(\sqrt{N_{\rm yr}/10})$.}

{While Gaussianity of the noise is an approximation, it allows us to capture the dependence on depth easily. Furthermore, relaxing it would require large amounts of simulations (mock galaxies injected in simulated or real images) in order to characterize the regimes where noise is non-Gaussian, the dependence on depth, and how it affects other statistics such as galaxy number counts.}

%% file: derivatives.tex
Several of the metrics defined \cwr{i}n this paper are based upon propagating the impact of variations in depth into variations in the mean redshift or surface density of a sample.  If the effects of depth variations in different bands are independent (i.e., cross-terms are negligible) and the differences are small, this can be expressed as a combination of partial derivatives and shifts in each band's depth:

\begin{equation}
   \Delta x = \frac{\partial{x}}{\partial {m_{5,{u}}}} \Delta{m_{5,{u}}} + \frac{\partial{x}}{\partial {m_{5,{g}}}} \Delta{m_{5,g}} + \frac{\partial{x}}{\partial {m_{5,{r}}}} \Delta{m_{5,{r}}} + \frac{\partial{x}}{\partial {m_{5,{i}}}} \Delta{m_{5,{i}}} + \frac{\partial{x}}{\partial {m_{5,{z}}}} \Delta{m_{5,{z}}} + \frac{\partial{x}}{\partial {m_{5,{y}}}} \Delta{m_{5,{y}}},
   \label{eq: derivatives}
\end{equation}
where $x$ is the quantity of interest (e.g., the surface density of objects in some redshift range, or the mean redshift of some sample), $m_{5,{\alpha}}$ is the 5-$\sigma$ point source depth in band $\alpha$, and $\Delta$ represents the difference (e.g., from the overall mean) in some quantity.  We have tested this assumption of independence by comparing the prediction to this formula to the result when the depths in all bands are changed simultaneously by the same amount; the results agree \cwr{within $ \lesssim10\%$} in all cases, implying that a more sophisticated analysis is not needed for our purposes.

As a result, if one precomputes the partial derivatives used in \autoref{eq: derivatives}, it is then possible to immediately combine them with the set of predicted depths at some location in the sky to obtain the change in the predicted density or mean redshift for some sample of interest in that region.  Similarly, one could compute the power spectrum or correlation of quantity $x$ by a simple propagation from the set of power spectra of $m_{5,\alpha}$.  

The calculation of the derivatives themselves is built upon the simulated datasets described in \autoref{sec:simulation}.  As was described in that section, these datasets were created by applying noise appropriate to the average 5-$\sigma$ point source depth  in each LSST \cwr{filter} after each survey year, along with additional scenarios where the depth in a single band or the depth in all bands simultaneously is shifted compared to the mean. However, the true (noiseless) magnitudes and true redshift of each object remain known.

Our analysis focuses on the sensitivity of either the mean true redshift or the total surface density for samples placed in each of six bins based upon their photometric redshifts (photo-$z$'s), evenly spaced from redshift zero to 1.2 (i.e., bins covering the ranges 0--0.2, 0.2--0.4, \cwr{and so} on up to 1.0-1.2), following \cite{DESCSRD}. Due to both the limited information contained in photometric measurements and the noisiness of the magnitude estimates for each object, the photo-$z$'s will scatter about the true redshifts of each object, with that scatter depending on the depth of the observations; greater errors will cause objects to shift from bin to bin, changing the densities and true mean redshift for each one. Additionally, noise in $i$ magnitudes will cause objects to scatter above or below the gold sample limiting magnitude,  $i_{\rm lim}$, for a given year; we also wish to capture sensitivity to this effect.

However, to assign objects to photometric redshift-based bins, we must compute photo-$z$ estimates for each object based on its degraded photometry.  For both simplicity and speed of computation, we utilize a Random Forest-based algorithm for this purpose (similar to that used in \citet{Zhou2023}).  We use the \texttt{scikit-learn} implementation of Random Forests with 
\texttt{n\_estimators = 100}, \texttt{max\_depth = 30}, and \texttt{max\_features = `sqrt'}.  For each year of the LSST survey, we separately train a random forest based upon the simulated set of objects used as inputs in \autoref{sec:simulation}, but independently degraded according to the expected 5$\sigma$ depths in that year (to avoid the possibility of overfitting).  The random forests are computed using $u-g$, $g-r$, $r-i$, $i-z$, $z-y$, and the $i$-band magnitude as features for predicting the true redshift of each object.

For a given simulation (corresponding to some survey year and set of depths in each band), we then compute photometric redshifts for every object using the random forest trained based on the typical depth for that year.  

As photometric errors increase or decrease, errors in the random forest redshift predictions will degrade or improve correspondingly, causing objects to scatter from redshift bin to redshift bin in varying ways.  Similarly, the number of objects that scatter above or below $i_{\rm lim}$ will change if the $i$-band depth does.  We therefore tabulate each bin's true mean redshift and the number of objects placed within it for each of the simulations described in \autoref{sec:simulation}.

We then wish to calculate the partial derivatives of redshift or density with respect to depth; i.e., either $\frac{\partial\langle z \rangle}{\partial m_{5,\alpha}}$ or $\frac{1}{n}\frac{\partial n}{\partial m_{5,\alpha}}$ (which we focus on rather than $\frac{\partial n}{\partial m_{5,\alpha}}$ as the fractional change is of greater interest).  We do this by performing a second-order polynomial fit to either $\langle z\rangle$ or $n / n(\Delta m_{5,\alpha} = 0)$ as a function of $m_{5,\alpha}$, using the tabulated results from each simulation.  The term of the linear coefficient in this fit corresponds to the derivative of interest, while the strength of the quadratic term can be used to assess how good the assumption that changes are linear in depth actually is.  We find that for all the years considered, the incorporation of a quadratic term has only small effects ($\lesssim 10\%$) for depth shifts of $<0.5$ mag, so our procedure should be sufficient \cwr{for} the needs of this analysis. 